\documentclass{elsart}
\usepackage{amssymb}
\usepackage[pctex32]{graphics}
\usepackage{graphicx}
\graphicspath{{converted_graphics/}}
\begin{document}
\newcommand{\be}{\begin{equation}}
\newcommand{\ee}{\end{equation}}
\newcommand{\ben}{\begin{eqnarray}}
\newcommand{\een}{\end{eqnarray}}
\newcommand{\nn}{\nonumber \\}
\newcommand{\ii}{\'{\i}}
\newcommand{\pp}{\prime}
\newcommand{\nd}{{\noindent}}
\begin{frontmatter}
\title{Statistical Complexity of \\ Sampled Chaotic Attractors}
\author[mdp,CONICET]{Luciana De Micco},
\ead{lucianadm55@gmail.com}
\author[mdp]{Juana Graciela Fern\'andez},
\ead{gfernand@fi.mdp.edu.ar}
\author[mdp,CONICET]{Hilda A. Larrondo \corauthref{cor}}
\ead{larrondo@fi.mdp.edu.ar}
\author[iflp,CONICET]{Angelo Plastino}
\ead{plastino@fisica.unlp.edu.ar} and
\author[brazil,calculo,CONICET]{Osvaldo A. Rosso}
\ead{oarosso@fibertel.com.ar, oarosso@gmail.com}
\corauth[cor]{Corresponding author. Phone: +54-223-481-6600}

\address[mdp]{Departamentos de F\'{i}sica y de Ingenier\'{i}a Electr\'{o}nica,\\
              Facultad de Ingenier\'{\i}a,Universidad Nacional de Mar del Plata.\\
              Av. Juan B. Justo 4302, 7600 Mar del Plata, Argentina}

\address[iflp]{Instituto de F\'isica, CCT-Conicet\\
             Universidad Nacional de La Plata (UNLP).\\
             C.C. 727, 1900 La Plata, Argentina.}

\address[calculo]{Chaos \& Biology Group,
                  Instituto de C\'alculo, \\
                  Facultad de Ciencias Exactas y Naturales,
                  Universidad de Buenos Aires.\\
                  Pabell\'on II, Ciudad Universitaria.\\
                  1428 Ciudad de Buenos Aires, Argentina.}

\address[brazil]{Departamento de F\'{\i}sica, Instituto de Ci\^encias Exatas,\\
                 Universidade Federal de Minas Gerais.\\
                 Av. Ant\^onio Carlos, 6627 - Campus Pampulha. \\
                 31270-901 Belo Horizonte - MG, Brazil.}

\address[CONICET]{Fellow of CONICET-Argentina}

\begin{abstract}
We analyze the statistical complexity vs. entropy
plane-representation of sampled chaotic attractors as a function
of the sampling period $\tau$.
It is shown that if the Bandt and Pompe procedure is used
to assign a probability distribution function (PDF) to the
pertinent time series, the statistical complexity measure (SCM)
attains a definite maximum for a specific sampling period $t_M$.
If the usual histogram approach is used instead in order to assign
the PDF to
the time series, the SCM remains almost constant at any sampling 
period $\tau$. The significance of $t_M$  is further investigated
by comparing it with typical times given in the literature for the
two main reconstruction processes: the Takens' one in a delay-time
embedding, and the exact Nyquist-Shannon reconstruction. It is
shown that $t_M$ is compatible with those times recommended as
adequate delay ones in Takens' reconstruction. The reported
results correspond to three representative chaotic systems having
correlation dimension $2<D_2<3$.
One  recent experiment confirm the analysis presented here.


\end{abstract}

\maketitle

\end{frontmatter}

{\bf Version: V2-15}

\section{Introduction}
\label{sec:Intro}

The study of  randomness started with
Poincar\'e, but it took almost $80$ years for digital electronics
to make enough computational resources available so that the
investigation on the transition from order to randomness could
lead to the fascinating issue of deterministic chaos. One hallmark
of chaotic systems is the presence of recognizable patterns in
their state space. Pattern-formation plays a fundamental role  in
the definition of i) \textsl{structural complexity} by Crutchfield
\cite{Crutchfield1989} and ii) a number of other complexity
measures, such as the so-called  statistical complexity (SCM),
relevant for this paper. For an excellent review on complexity
measures see \cite{Wackerbauer1994}. The SCM, based on the notion
of \textsl{disequilibrium} in Statistical Space, was proposed by
L\'opez Ruiz, Mancini and Calbet \cite{Lopez1995} and improved
upon later in several papers
\cite{Martin2003,Lamberti2004,Rosso2007A}. The SCM version
employed here was advanced in \cite{Lamberti2004}. In a related
vein, the importance of using a ``\textsl{causal}" Probability
Distribution Function (PDF) to analyze the stochasticity-degree of
chaotic and pseudo stochastic systems was emphasized in
\cite{Rosso2007A} and \cite{Rosso2007B,DeMicco2008,DeMicco2009}.

In this field of endeavor it  is of importance to study just how
the sampling period $\tau$-influences the final results given that
 digital instrumentation is widely used  in
concomitant experiments. The most common sampling criterion is
given by the sampling theorem by Nyquist and Shannon
\cite{Nyquist1928,Shannon1949}. It stipulates  that the
\textsl{exact reconstruction} of a continuous signal is possible
whenever its Fourier spectrum vanishes for frequencies above a
certain $B=f_{max}$ (the bandwidth). An \textit{infinite} number
of regularly sampled values is required and $\tau$ must satisfy
the inequality $\tau<t_{NS}=1/(2B)$ where $t_{NS}$ is the
Nyquist-Shannon minimum sampling interval. The exact original
continuous signal is recovered from the time series by using an
\textit{ideal low pass filter}. Note that in actual scenarios the
number $N$ of samples is always finite and an exact reconstruction
is not possible.

In a different vein, Takens \cite{Takens1981} demonstrated that
chaotic  attractors generated by dynamical systems may be
\textsl{reconstructed} if one has access to the \textit{infinite}
time series of regularly sampled values for just one state
variable. Takens' procedure allows one to obtain the main
parameters characterizing the chaotic system, like  Lyapunov
exponents, dimensions, etc. For an impressive web-site containing
very useful routines for nonlinear time series analysis see
\cite{Hegger1999}. The ensuing reconstruction procedure uses
``delayed-samples"  with
$t_T$ as the delay time, to generate a
$d$-dimensional vector (in a $d$-dimensional embedding space).
Note that the time delay $t_T$ must be a multiple of the   %
sampling period $\tau$, since we only possess data gathered at
those times.
Let us remark  that in the context of Takens' theorem the vocable  %
\textsl{reconstruction} adopts  a different meaning than that
assigned by the Nyquist-Shannon theorem. In this paper we
differentiate  between a \textsl{Takens reconstruction} and a
\textsl{Nyquist reconstruction}. {\it We will carefully  look  at
just how the representative point of a chaotic attractor in an
entropy-complexity plane ($H \times C$) changes with the sampling
period}. It will be shown that when the \textsl{causal PDF}
prescribed by the Bandt and Pompe procedure \cite{Pompe2002} is
used, the ensuing complexity $C=C^{(BP)}$ exhibits a maximum for a
given sampling period $\tau=t_M$. However, such maximum does not
appear if the
 conventional  histogram-approach is used instead in assigning a
PDF to the time series and thus a $C=C^{(hist)}$ is used. We call
the later histogram-distribution  a \textsl{non-causal PDF}. In
point of fact, that of causality constitutes a difficult topic
(see for instance \cite{Pearl2009}). In this paper a
\textsl{causal PDF} is %
 one that takes into account the temporal correlation between successive samples, which
entails using a PDF with a symbol assigned to each trajectory's
piece of  length
$L=d \cdot \tau$.%

It is %
of interest, for both the Takens
and the Nyquist-Shannon reconstruction processes, to analyze the
relation between
{\it i)\/} $t_M$ and
{\it ii)\/} other related temporal quantities proposed in the literature such as $t_T$ or $t_{NS}$.
This is one of the topics to be here  investigated. In the case of Takens' reconstruction
it is well known that for a finite number of samples  $N$ the %
quality of the reconstruction requires the selection of an
adequate value for $t_T$.
There exist  in the literature several different recommendations
for a ``good"   $t_T $, like the first zero of the autocorrelation
function, the first minimum of the mutual information function,
etc. \cite{Fraser1986}. For an interesting discussion see
\cite{Crutchfield1987,Kantz1999}. In the case of Nyquist's
reconstruction, a bandwidth criterium must be chosen in order to
evaluate the Nyquist-Shannon time $\tau=t_{NS}$ (because the
spectrum of chaotic systems is not band-limited).

In Section \ref{sec:takens} we review the Takens embedding theorem
and the delay embedding reconstruction process. Section
\ref{sec:nyquist} deals with the Nyquist-Shannon reconstruction
process. In Section \ref{sec:CBP} the specific SCM used in the
paper is reviewed. Results for three paradigmatic examples:
{\it (a)\/}  the Lorenz Chaotic Attractor;
{\it (b)\/}  the R\"ossler Chaotic Attractor and
{\it (c)\/} the chaotic attractor named ${\mathbf B}_7$ by Chlouverakis and Sprott \cite{Chlouverakis2005} are
reported in Section \ref{sec:results}. The first two attractors
have been extensively studied in the literature but they both have
a correlation dimension $D_2\approx2$ while the attractor
${\mathbf B}_7$ has $D_2\approx 2.719$, covering almost all the
$3D$ state-space. Conclusions and remarks are included in Section
\ref{sec:Conclusions}.  It is interesting to remark that the
analysis performed here has been experimentally verified
\cite{Soriano2011}.

\section{Takens' embedding theorem}
\label{sec:takens}

Let $d\mathbf{x}/dt=\mathbf{f}(\mathbf{x})$  be
an $m$-dimensional continuous dynamical system.  A scalar
measurement $s(t) \equiv s(\mathbf{x}(t))$ is a projection of the
state $\mathbf{x}$ onto an interval $I \in \mathbb{R}$. The goal
of all the embedding theorems proposed in the literature is to
obtain a $d$-dimensional \emph{embedding space} $\mathfrak{A}$ in
which $\mathbf{x}(t)$ can be reconstructed by using only $s(t)$.
This \textsl{reconstruction} does not need to be exact, especially
in the case of chaotic dissipative systems that usually have
attractors with box counting dimension $D_{BC}$ smaller than $m$,
the dimension of the state space. In this context,
\textsl{reconstruction} merely indicates  that
$\mathbf{\widetilde{x}}(t) \in \mathfrak{A}$ shares with
$\mathbf{x}(t)$ some characteristics. The main requirements a
reconstructed space must fulfill are:
\begin{enumerate}
\item[\textit{(a)}] Uniqueness of the dynamics in the reconstructed space.
\item[\textit{(b)}] The reconstructed  attractor must have dimensions, Lyapunov exponents and entropies
identical to those of the original attractor.
\end{enumerate}

Consequently,  the embedding of a compact smooth manifold $A$ into
$\mathbb{R}^d$ is defined to be a one-to-one $\mathfrak{C}^1$ map
$\mathcal{F}$, with a Jacobian $DF(\mathbf{x})$ which has full rank
everywhere. Let us assume that an infinite length-time scalar
series $\{s_n ; n=1,2, \cdots, \infty \}$ is obtained by measuring
one component of the $m$-dimensional vector field at evenly spaced
times $t_n = n\cdot \tau$, with $\tau$ the sampling period. A
time-delayed vector field is constructed as follows:
\begin{equation}
\label{eq:delayVector}
{\widetilde{x}}^{(1)}_n=s_n~; \;
{\widetilde{x}}^{(2)}_n=s_{n+1}~; \; \cdots  ~; \;
{\widetilde{x}}^{(d)}_{n}=s_{n+d-1} \ .
\end{equation}

Takens  proved \cite{Takens1981}  that time-delay maps of
dimension $2d+1$ have the generic property of being the embedding
of a compact manifold with dimension $d$, if:
(1) the measurement function $s:A\rightarrow \mathbb{R}$ is $\mathfrak{C}^2$ and
(2) either the dynamics or the measurement couples all degrees of
freedom. In the original version by Takens, $d$ is the integer
dimension of a smooth manifold, the phase space containing the
attractor.
Thus $d$ can be much larger than the attractor
dimension. Sauer {\it et al.\/} \cite{Sauer1993} were able to
generalize the theorem into what they call the Fractal Delay
Embedding Prevalence Theorem. Let $D_{BC}$ be now the box counting
dimension of the (fractal) attractor. Then, for almost every
smooth measurement function $s$ and any sampling time $\tau>0$,
the delay map into $\mathbb{R}^d$ with $d > 2 D_{BC}$ is an
embedding if: (1) there are no periodic orbits of the system with
period $\tau$ or $2 \tau$ and (2) there only exists a finite
number of periodic orbits with period $p \tau$, with $p>2$. Thus
the main result of the embedding theorems is that it is not the
dimension $m$ of the underlying state space what is important for
ascertaining the minimal dimension of the embedding space, but
only the fractal dimension $D_{BC}$ of the support of the
invariant measure generated by the dynamics in the state space. In
dissipative systems $D_{BC}$ can be much smaller than $m$. Let us
further remark that in favorable cases an attractor might  be
reconstructed in spaces of dimension $d$ such that $D_{BC} \le d \le 2D_{BC}$.
For example, for the determination of the
correlation dimension, events of measure zero can be neglected;
thus any embedding with a dimension larger that $D_{BC}$ is
sufficient.

From a mathematical point of view,  and for an {\it infinite}
number of data items (known with infinite precision as assumed in
embedding theorems),    %
 the time-delay $t_T$ is an arbitrary multiple of $\tau$.
Thus, there exists no rigorous way of determining  $t_T$'s
optimal value. But in a real scenario with a {\it finite number}
 $M$ of data items, the specific value adopted by the time delay is
quite important. Moreover, it is even unclear what properties the
optimal value should have for
best estimating  a continuous system's specific property.  %
Many different methods have been suggested to
estimate the time-delay
\cite{Hegger1999,Kantz1999,Casdagli1991,Gibson1992}. In the case
of  Takens' reconstruction procedure it is possible to use a time delay
$t_T > \tau$ as pointed out above.
Since in this paper the sample period $\tau$ will be movable we will consider that both times are equal
($t_T= \tau$).

The analysis of
{\it (i)\/} linear autocorrelations  and
{\it (ii)\/} average mutual information for a time series are two of the criteria often
used in the
 literature to determine the best $t_T$-region. Several
characteristic times have been recommended using these two types
of analysis. Let us recapitulate.
\begin{enumerate}
  \item {\it Time-delays induced by the discrete linear autocorrelation function.\/}
        Let ${\mathcal S} \equiv \{s(n) ; n=1, \cdots, M \}$ be the measured component of  the vector field
        (the time series).
        The discrete linear autocorrelation function is a vector $R_i$ defined as:
\begin{equation}
\label{eq:autocorrx}
        R_i~=~\frac{1}{M}~\sum_{n=0}^{M-i-1} [s({n+i}) - <s>] \cdot [s(n) -<s>] \ ,
\end{equation}
        with $<s> = \sum_{i=1}^{M} s_i$  the mean value for the time series. \\

Four characteristic time delays are considered here:
   \begin{enumerate}
      \item [\textit{(i)}]The first zero crossing of $R_i$ (if it exists). 
         Let $\tau$ and $i_0$ be,
            respectively, the sampling period and the first zero crossing of $R_i$.
            The characteristic time is $t_0~=~i_0~\tau$.
      \item [\textit{(ii)}]The first zero crossing of $R'_i~=~R_{i+1}-R{_i}$ gives the first minimum of $R$.
            We call this characteristic time $t_1=i_1~\tau$.
      \item [\textit{(iii)}] The first zero crossing of $R''_i~=~R'_{i+1}-R'_{i}$ yields the first curvature-change of $R$.
            We call this characteristic time $t_2=i_2~\tau$.
      \item [\textit{(iv)}]Let  $i=i_{1/e}$ be the smallest $i$ making $R_i$ to decay to less than $R_0 /e$.
         The corresponding characteristic time is $t_{1/e}=i_{1/e}~\tau$. \\
   \end{enumerate}

   \item {\it Delay-time induced by the discrete  Mutual Information function.\/}
         Let ${\mathcal S} \equiv \{s(n) ; n=1, \cdots, M \}$ be the measured component of the vector field (the time series).
         The discrete Mutual Information function is a vector  $I_i$, defined as:
\begin{equation}
\label{eq:infomutua}
        I_i=~-\sum_{k,l} p_{kl}(s(n),s(n+i)) \ln \left[ \frac{p_{k,l}(s(n),s(n+i))}{p_k(s(n)) \cdot p_l(s(n+i))} \right] \ .
\end{equation}
\end{enumerate}

Equation (\ref{eq:infomutua}) is determined as follows:
{\it (a)\/} the real interval $[a,b]$  covered by the time series is
partitioned into  $N_{box}$ subintervals, equal sized consecutive non
overlapping subintervals;
{\it (b)\/}  $p_{k}$  is the probability to find a time series' value, $s(n)$, in the $k$-th  interval,
and $p_{k,l}$ is the joint probability for simultaneously encountering a time series' value, $s(n)$,
in the $k$-th interval while the time series' value found at the $i$-th posterior positions, $s(n+i)$,
falls into the the $l$-th interval.
This quantity can be quite easily computed, for sufficiently small sizes of the partition elements
(sufficiently high values of $N_{box}$), and, provided the
attractor dimension is $\le 2$,
this expression has no systematic dependence on $N_{box}$ \cite{Kantz1999,Abarbanel1996}.

There exist good arguments for asserting that, if the time-delayed mutual
information exhibits a marked minimum for a certain value of the
delay $i_{I}$, then this is a good candidate for a ``reasonable"
time delay. The corresponding characteristic time is given by
$t_{I}~=~i_{I}~\tau$.
Nevertheless when one finds that the minimum of the mutual
information $t_{I}$ lies at considerably larger times than
$t_{1/e}$ (the  decay of autocorrelation function), it is worth
optimizing the time-lag inside this range \cite{Kantz1999}.

%
\section{Nyquist-Shannon theorem and the minimum sampling time}
\label{sec:nyquist}

The well known Nyquist-Shannon sampling theorem
\cite{Nyquist1928,Shannon1949} is based on the Fourier Transform (FT).
It states that a function $s(t)$, $t \in [-\infty,\infty]$,
containing no frequencies higher than $B$ in its FT,
is completely determined by giving its coordinates at an infinite
series of points (samples)  spaced $\tau~<~t_{NS}=1/(2B)$  apart.

Consider the case of chaotic signals that are not band-limited.
A value of $B$ may be defined from the
Discrete FT as follows. Let $\{S\}$,  be the Discrete FT (DFT) of $\{s\}$  defined by
\begin{equation}
\label{eq:fourier}
S_k~=~\sum_{j=1}^{M} s_j~e^{-i\;2\pi(j-1)(k-1)/M} \ .
\end{equation}
The power contained in the first $i$ frequency components of the DFT's is given by
\begin{equation}
\label{eq:parpow}
PW_i=\sum_{k=1}^{i} S_k ~S^*_k,
\end{equation}
where $^*$ refers to  complex conjugation  while
the index $i\leq M$. The full power $PW$  is given by Eq. (\ref{eq:parpow}) with $i=M$.
To define $B$ we choose  the value
$i=i_\alpha$ in such a way as to make $PW_{i_\alpha}=\alpha PW$.
The maximum frequency $B$ for such $\alpha-$value is then defined
as $B=i_\alpha /\tau$. Once this maximum frequency is obtained,
the Nyquist-Shannon criterion prescribes that
$t_{NS}^{(\alpha)}=1/(2B)$.
Our present calculations were made with $0.80\leq\alpha\leq0.99$.

\section{The Statistical Complexity Measure $C$ using Bandt and Pompe's prescription}
\label{sec:CBP}

The SCM $C$ is  an informational quantifier. As
such, it is a functional of a probability distribution function
(PDF). In this case we refer to an appropriate PDF associated with
a \textit{time series}. Given the PDF, $P \equiv \{p_i ; i=1,\cdots, N\}$, there are several manners to obtain the
functional and a full discussion of the subject would exceed the
scope of this presentation (for a comparison amongst different
complexity measures see the excellent paper by Wackerbauer
\textit{et al.} \cite{Wackerbauer1994}).

In the present work we adopt for the SCM  the functional form
introduced in L\'opez Ruiz-Mancini-Calbet seminal paper
\cite{Lopez1995} with the modifications advanced by Lamberti
{\it et al.} \cite{Lamberti2004}. This functional form is an intensive (in a
thermodynamics sense) statistical complexity $C[P]$ given by
\begin{equation}
\label{eq:inten}
C[{P}]=Q_{J}[{P,P_e}]\cdot H[{P}] \ ,
\end{equation}
where $H$ denotes the amount of ``disorder" given by the normalized Shannon entropy
\begin{equation}\label{eq:sha}
H[P] = S[P]/S_{max},
\end{equation}
and $Q_{J}$ is  the so-called ``disequilibrium", defined in terms
of the extensive Jensen-Shannon divergence (which induces a
squared metric) \cite{Lamberti2004}:
\begin{equation}
\label{eq:disequi}
Q_{J}[{P,P_e}] = Q_0 \cdot J[P,P_e] =  Q_0 \cdot \{S[(P+P_e)/2]-S[P]/2-S[P_e]/2 \} \ .
\end{equation}

In these equations $P_e =\{ 1/N, \cdots,1/N\}$ stands for the {\it
uniform} distribution,  while $S[P]=-\sum _{j=1}^{N}~p_j~\ln( p_j)$
is the Shannon entropy corresponding to the PDF $P$. We use two
normalization constants, namely,
\begin{equation}
\label{eq:Smax}
S_{max}= S[P_e] = \ln N \ ,
\end{equation}
and
\begin{equation}
\label{eq:q0j}
Q_0=-2 \left\{ \left( \frac{N+1}{N} \right) \ln(N+1) - \ln (2N) +  \ln N \right\}^{-1} \  .
\end{equation}

The normalization constant $Q_0$ is equal to the inverse of the
maximum possible value of $J[P, P_e]$. This value is obtained when
one component of $P$, say $p_m$, is equal to one and all the
remaining $p_i$'s are equal to zero. We have then $0 \le H \le 1$
and $0 \le Q_{J} \le 1$.

The disequilibrium $Q_J$ is an  intensive thermodynamical quantity
that reflects on the systems' \textsl{architecture}, being
different from zero only if there exist \textsl{privileged}, or
\textsl{more likely} states among the accessible ones. $C[P]$
quantifies the presence of correlational structures as well
\cite{Martin2003,Lamberti2004}. The opposite extremes of perfect
order and maximal randomness possess no structure to speak of and,
as a consequence, their $C[P]=0$. In between these two special
instances a wide range of possible degrees of physical structure
exist, degrees that should be reflected in the features of the
underlying probability distribution.

The complexity measure constructed in this way is intensive, as
many thermodynamic quantities \cite{Lamberti2004}. We stress the
fact that the above SCM is not a trivial function of the entropy
because it depends on two different probabilities distributions,
the one associated to the system under analysis, $P$, and the
uniform distribution, $P_e$. Furthermore, it was shown that for a
given $H$ value, there exists a range of possible SCM values
\cite{Martin2006}. Thus, it is clear that $C$ carries important
additional information (related to the correlational structure
between the components of the physical system) that is not
contained in the entropic functional.

A detailed analysis of the $C$-behavior
demonstrates the existence of bounds to $C$ that we called
$C_{max}$ and $C_{min}$. These bounds can be systematically
evaluated by recourse to a careful geometric analysis performed in
the space of probabilities $\Omega$ \cite{Martin2006}. The
corresponding values of  $C_{max}$ and $C_{min}$  depend only on
the probability space's dimension and, of course, on the
functional form adopted by the amount of disorder $H$ and the
disequilibrium $Q$.  $H \times C$ diagrams are important and yield
system's information independently of the values that the
different control parameters may adopt. The bounds  also provide
us with relevant information that depends on the system's
particular characteristics,
as for instance, the existence of global extrema, or the peculiarities of the system's configuration%

As pointed out above, the  PDF $P$ itself is not a uniquely
defined object and several approaches have been employed in the
literature to ``associate" $P$ to a given time series.  Just to
mention some frequently used $P-$extraction (from the time series)
procedures, one has:
{\it a)\/} frequency counting \cite{Rosso2009C}, {\it b)\/} time
series histograms \cite{DeMicco2008}, {\it c)\/} binary
symbolic-dynamics \cite{Mischaikow1999}, {\it d)\/} Fourier
analysis \cite{Powell1979}, {\it e)\/} wavelet transforms
\cite{Blanco1998,Rosso2001}, {\it f)\/} partition entropies
\cite{Ebeling2001}, {\it g)\/} discrete entropies
\cite{Amigo2007}, {\it h)\/} permutation (Bandt-Pompe) entropies
\cite{Pompe2002,Keller2005}, among others. There is  ample liberty
to choose among them and the specific application must be
carefully analyzed so as
to make a good choice. Rosso {\it et al.}  \cite{Rosso2007A}
showed that the last mentioned methodology may be profitably used
in the plane $H \times C$ so as to separate and differentiate
amongst stochastic, chaotic, and deterministic systems.  It was
shown in \cite{Rosso2009A,Rosso2009B}  that temporal correlations
are nicely displayed by the Bandt and Pompe PDF.

Summing up,  different \textsl{symbolic sequences} may be assigned
to a given time series. If one symbol $a $ of the finite alphabet
$\mathfrak{A}$ is assigned to each $x_t$ of the time series,
the \textsl{symbolic sequence} can be regarded as a \textsl{non causal coarse grained }
description of the time series because the 
resulting PDF will not have any \textsl{causal information}. The usual
histogram-technique corresponds to this kind of assignment. For
extracting $P$ via an histogram one divides the interval $[a,b]$
(with $a$ and $b$ the minimum and maximum values in the time
series) into a finite number $N_{bin}$ of non overlapping equal
sized consecutive subintervals $A_i:[a,b]=\bigcup_{i=1}^{N_{bin}}
A_i$ and $A_i\bigcap A_j=\emptyset~\forall i\neq j$.

Note that $N$ in equations (\ref{eq:Smax}) and (\ref{eq:q0j}) is equal to $N_{bin}$.
Of course, in this approach the temporal order of the time-series plays no role
at all. The quantifiers obtained via the ensuing PDF-histogram are called in
this paper, respectively, $H^{(hist)}$ and $C^{(hist)}$. Let us
also point out that for time series with a finite alphabet it is
relevant to consider a judiciously chosen optimal value for
$N_{bin}$ (see e.g. De Micco {\it et al.} \cite{DeMicco2008}).

{\it Causal information\/}  may be duly incorporated into the
construction-process that yields $P$ if %
one symbol of a finite alphabet $\mathfrak{A}$ is assigned to a
trajectory's portion, i.e.,  we assign ``words" to each
trajectory-portion.
The Bandt and Pompe methodology for extraction of the PDF
corresponds to this type of assignment and the resulting
probability distribution $P$  is thus a \textsl{causal coarse
grained} description
of the system. %
Note that, in the Bandt and Pompe approach a sort of  {\it coarse graining\/} and
{\it word construction\/} is effected (for methodological detail
see below).
 Note that there are other ways to get a causal PDF. The advantage of the Bandt and Pompe approach
lies in the fact  that it solves the problem of finding generation
partitions.

Thus one expect that for increasing patterns' length (embedding
dimension)  the Bandt and Pompe approach retains all relevant
essentials of the original continuous (in space) dynamics.
The quantifiers obtained by appeal to this PDF are denoted in this paper as $H^{(BP)}$ and
$C^{(BP)}$, respectively.
A notable Bandt and Pompe result consists in yielding a clear improvement on the quality of
Information Theory-based quantifiers
\cite{Rosso2007A,Rosso2007B,DeMicco2008,DeMicco2009,Rosso2009A,Rosso2009B,Larrondo2005,Larrondo2006,Kowalski2007,Zunino2007,Rosso2008,Zunino2008A,Zunino2008B,Zunino2009,Zunino2010,Zunino2010b,Rosso2010A,Rosso2010B,Saco2010}.

In summarizing now the approach, note that Bandt and Pompe \cite{Pompe2002} introduced a simple and robust method to
evaluate the probability distribution taking into account the time
causality of the system dynamics. They suggested that the symbol
sequence should arise naturally from the time series, without any
model-based assumptions. Thus, they took partitions by comparing
the order of neighboring values rather than partitioning the
amplitude into different levels. That is, given a time series
 ${\mathcal S} = \{ x_t ; t = 1, \cdots , M \} $, an embedding
dimension $d > 1 ~(d \in {\mathbb N})$,  and an embedding delay
$T~(T \in {\mathbb N})$, the ordinal pattern of order $d$
generated by
\begin{equation}
\label{eq:vectores}
s \mapsto \left(x_{s-(d-1)T},x_{s-(d-2)T},\cdots,x_{s-T},x_{s}\right) \ ,
\end{equation}
is to be considered.
To each time $s$ we assign a $d$-dimensional vector that results from the evaluation of the
time series at times $s - (d - 1) T, \cdots , s - T, s$.
Clearly, the higher the value of $d$, the more information about the past is
incorporated into the ensuing vectors.
By the ordinal pattern of order $d$ related to the time $s$ we mean the permutation
$\pi = (r_0, r_1, \cdots , r_{d-1})$ of $(0, 1, \cdots ,d - 1)$ defined by
\begin{equation}
\label{eq:permuta}
x_{s-r_{d-1} T} \le  x_{s-r_{d-2} T} \le \cdots \le x_{s-r_{1} T}\le x_{s-r_0 T} \ .
\end{equation}
In this way the vector defined by Eq. (\ref{eq:vectores}) is converted into a unique symbol $\pi$.
In order to get a unique result we consider that $r_i < r_{i-1}$ if $x_{s-r_{i} T} = x_{s-r_{i-1} T}$.
This is justified if the values of ${x_t}$ have a continuous distribution so that equal values are
very unusual.

For all the $d!$ possible orderings (permutations) $\pi_i$ when the embedding dimension is $d$,
their associated relative frequencies can be naturally computed by the number of times this
particular order sequence is found in the time series divided by the total number of sequences,
\begin{equation}
\label{eq:frequ}
p(\pi_i)~=~ \frac{\sharp \{s|s\leq M-d+1; (s) \quad \texttt{has type}~\pi_i \}}{M-d+1} \ .
\end{equation}
In the last expression the symbol $\sharp$ stands for ``number".
Thus, an ordinal pattern probability distribution $P = \{ p(\pi_i), i = 1, \cdots, d! \}$
is obtained from the time series.

It is clear that this ordinal time-series' analysis entails losing
some details of the original amplitude-information.
Nevertheless,  a meaningful reduction of the complex systems to their
basic intrinsic structure is provided. Symbolizing time series, on
the basis of  a comparison of consecutive points allows for an
accurate empirical reconstruction of the underlying phase-space of
chaotic time-series affected by weak (observational and dynamical)
noise \cite{Pompe2002}. Furthermore, the ordinal-pattern
probability distribution is invariant with respect to nonlinear
monotonous transformations. Thus, nonlinear drifts or scalings
artificially introduced by a measurement device do not modify the
quantifiers' estimations, a relevant property for the analysis of
experimental data. These  advantages make the BP approach more
convenient than conventional methods based on range partitioning.
Additional advantages of the Bandt and Pompe method reside in its
simplicity (we need  few parameters: the pattern length/embedding
dimension $d$ and the embedding time lag $T$) and the extremely
fast nature of the pertinent calculation-process
\cite{Keller2005,Keller2003}. We stress that the Bandt and Pompe's
methodology is not restricted to time series representative of low
dimensional dynamical systems but can be applied to any type of
time series (regular, chaotic, noisy, or reality based), with a
very weak stationary assumption (for $k =d $, the probability for
$x_t < x_{t+k}$ should not depend on $t$ \cite{Pompe2002}).

The probability distribution $P$ is obtained once we fix the
embedding dimension $d$ and the embedding delay $T$. The former
parameter plays an important role for the evaluation of the
appropriate probability distribution, since $d$ determines the
number of accessible states, given by $d!$. Moreover, it was
established that the length $M$ of the time series must satisfy
the condition $M \gg d!$ in order to achieve a proper
differentiation between stochastic and deterministic dynamics
\cite{Rosso2007A}. With respect to the selection of the
parameters, Bandt and Pompe suggest in their cornerstone paper
\cite{Pompe2002}  to work with $3\leq d \leq 7$ with a time lag $T
= 1$ . This is what we do here (in the present work $d=6$ and
$T=1$ are used). Of course it is also assumed that enough data are
available for a correct attractor-reconstruction.

The time-causal nature of the Bandt and Pompe PDF allows for its
success in separating chaotic from stochastic systems in different
regions of the plane $H^{(BP)}\times C^{(BP)}$ \cite{Rosso2007A}.
The main properties of the statistical complexity to be here
employed are: \textit{(i)} it is able to grasp essential details
of the dynamics, because it employs a causal PDF,
\textit{(ii)} it is an intensive quantity (in the thermodynamical sense) and,
\textit{(iii)} it is capable of discerning both among different degrees of
periodicity and of chaos \cite{Rosso2007A}.
 Note that $H^{(BP)}$ (the normalized Shannon permutation entropy)
and $C^{(BP)}$ (the permutation statistical complexity measure),
like other Information Theory measures (i.e. relative entropy,
mutual information, etc.) are not system-invariant as fractal
dimensions or Lyapunov exponents are. However, these quantifiers
provide important and valuable information on the dynamical system
under analysis. For example, $H^{(BP)}$ can be consider a good
approximation to the Kolmogorov-Sinai entropy \cite{Pompe2002}, a
system invariant quantity. That is, even if different time series
generated by a nonlinear dynamical system are used for its
evaluation, the global behavior of the system will be captured and
characterized,
independently of the  particular time series chosen.
%
\section{Results}
\label{sec:results}

\subsection{Preliminaries}
\label{sec:Preliminaries}
Our results refer to three paradigmatic chaotic systems with a $3$-dimensional state space,
namely,
\begin{itemize}
\item The Rossler chaotic attractor, given by
\begin{equation}
\label{eq:rossler} \left\{
     \begin{array}{l}
                \dot{x}~=~b~+~x~(y-c) \\
        \dot{y}~=~-x~-~z \\
            \dot{z}~=~y~+~a~z \\
     \end{array}
\right. \ ,
\end{equation}
where the parameters used here are $a=0.45$, $b=2$, and $c=4$,
corresponding  to a chaotic attractor.\\

\item The Lorenz chaotic attractor given by
\begin{equation}
\label{eq:lorenz} \left\{
     \begin{array}{l}
        \dot{x}~=~\sigma(y ~-~x) \\
        \dot{y}~=~rx~-~y~-~xz \\
        \dot{z}~=~xy~-~bz
     \end{array}
\right. \ ,
\end{equation}
where the pertinent parameters are $\sigma=16$, $b=4$, and
$r =45.92$, corresponding to a chaotic attractor.\\

\item The chaotic attractor ${\mathbf B}_7$ of \cite{Chlouverakis2005}:
\begin{equation}
\label{eq:sprott} \left\{
     \begin{array}{l}
        \dot{x}~=~K+z~(x ~-~\alpha ~y) \\
        \dot{y}~=~z~(\alpha~x~-~\epsilon~y)\\
        \dot{z}~=~1~-~x^1~-~y^2
     \end{array}
\right. \ ,
\end{equation}
where $K=0.5$, $\alpha=7.0$ and $\epsilon=0.23$.
\end{itemize}

\begin{figure}
\newpage
\includegraphics[ width=0.7\textwidth]{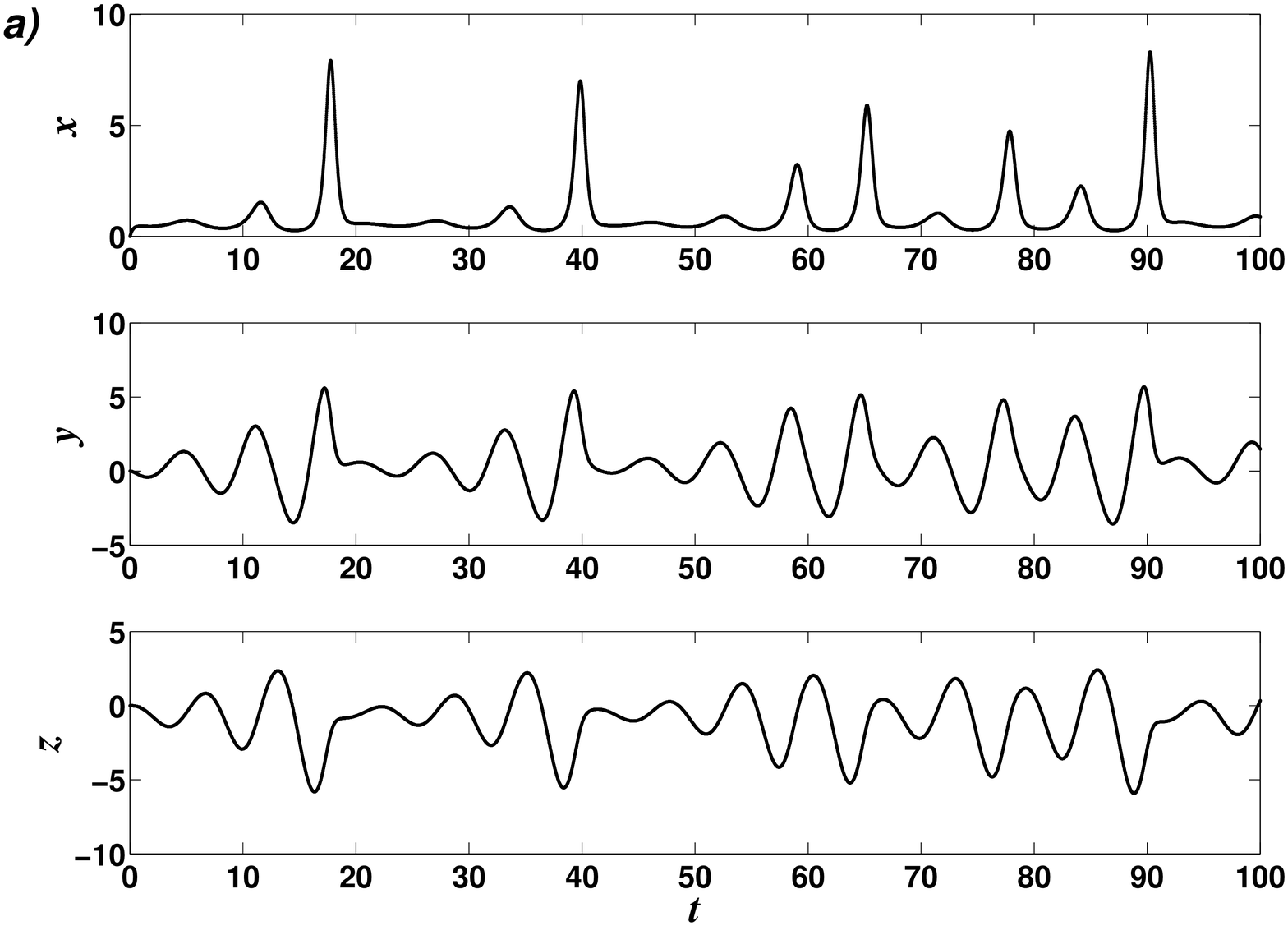}
\includegraphics[ width=0.7\textwidth]{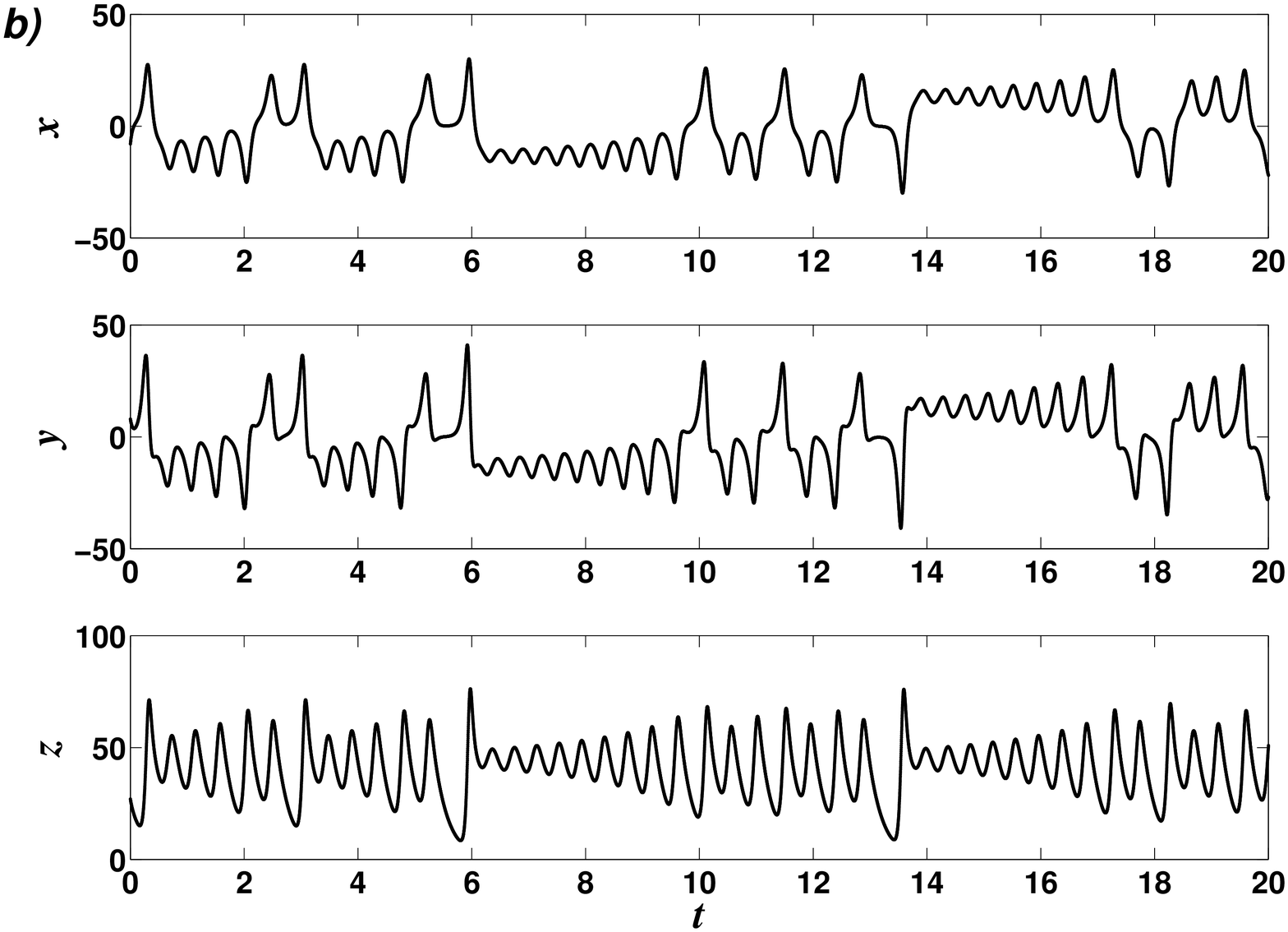}
\includegraphics[ width=0.7\textwidth]{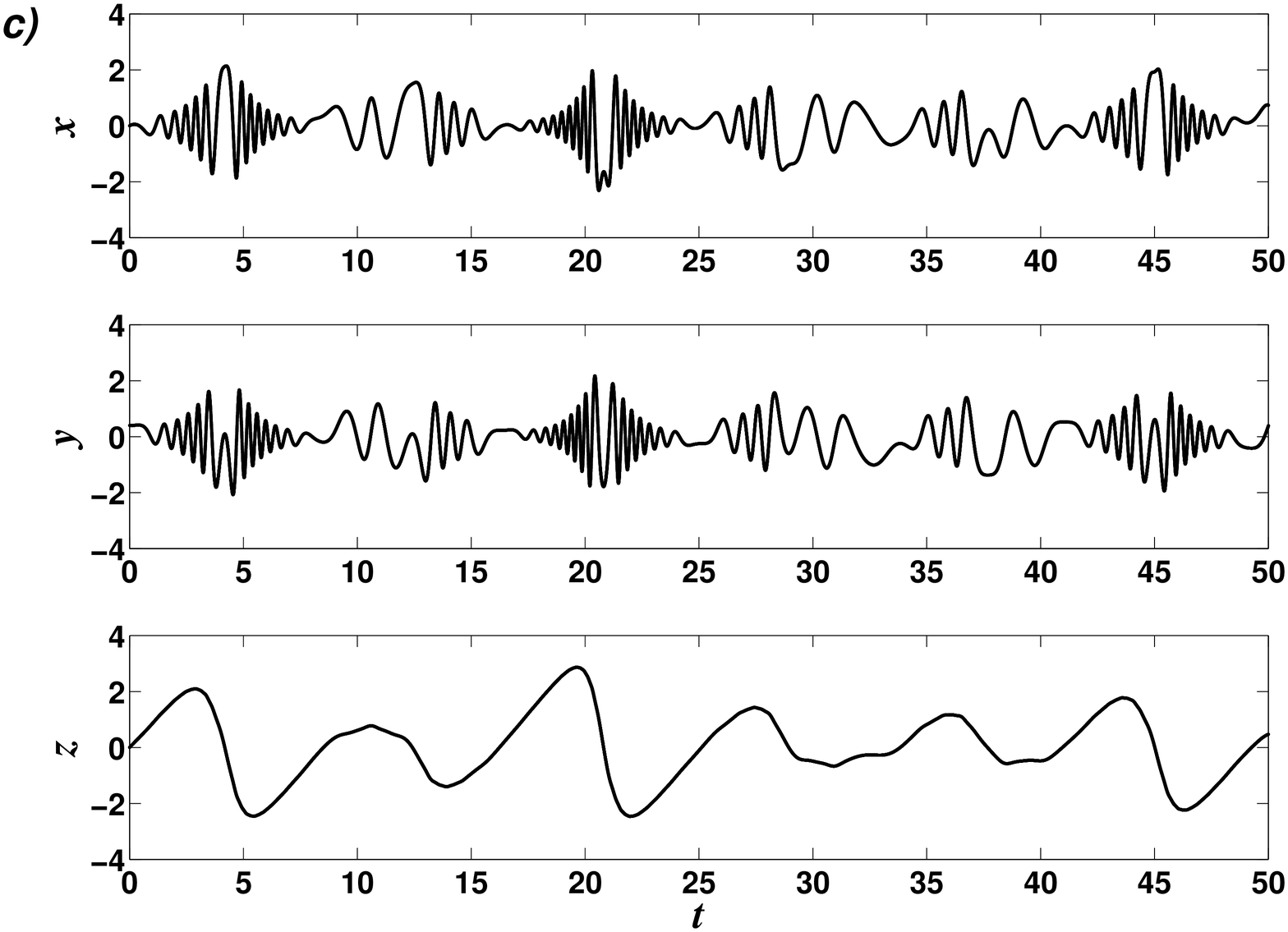}

\caption{
Evolution of the three coordinates as functions of time:
{\it (a)\/} Rossler variables $x$, $y$, and $z$ for $\tau =0.01$  with $a=0.45$, $b=2$, and $c=4$. 
{\it (b)\/} Lorenz variables  $x$, $y$,  and $z$ for $ \tau =0.001$ with $\sigma= 16$, $r=45.92 $, and $b=4$.
{\it (c)\/} ${\mathbf B}_7$ variables $x$, $y$,  and $z$ for $ \tau =0.001$
with $K~=~0.5$, $\alpha~=~7.0$, and $\epsilon~=~0.23$. }
\label{fig:xyzt}
\end{figure}

\begin{figure}
\newpage
\includegraphics[ width=0.7\textwidth]{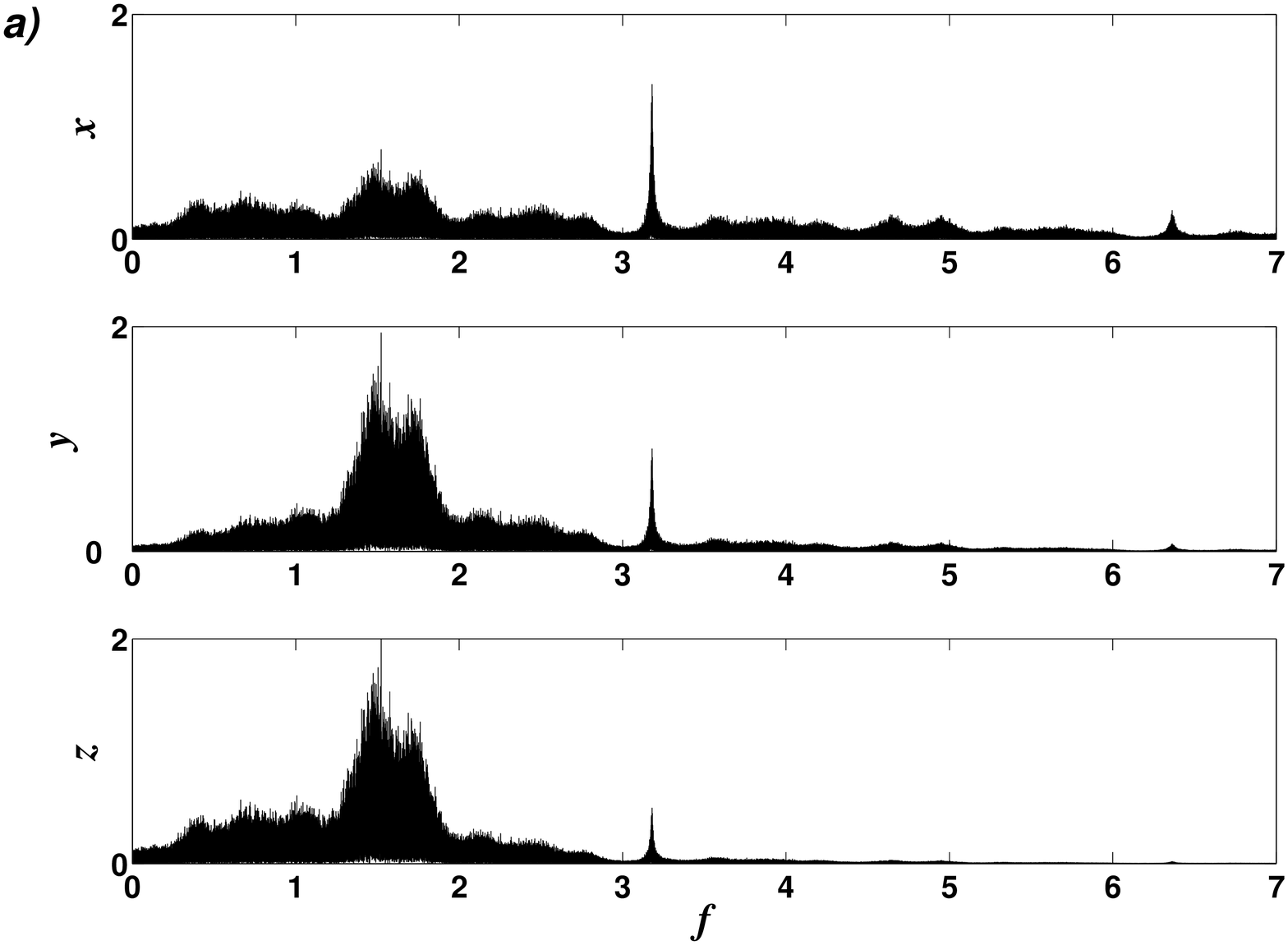}
\includegraphics[ width=0.7\textwidth]{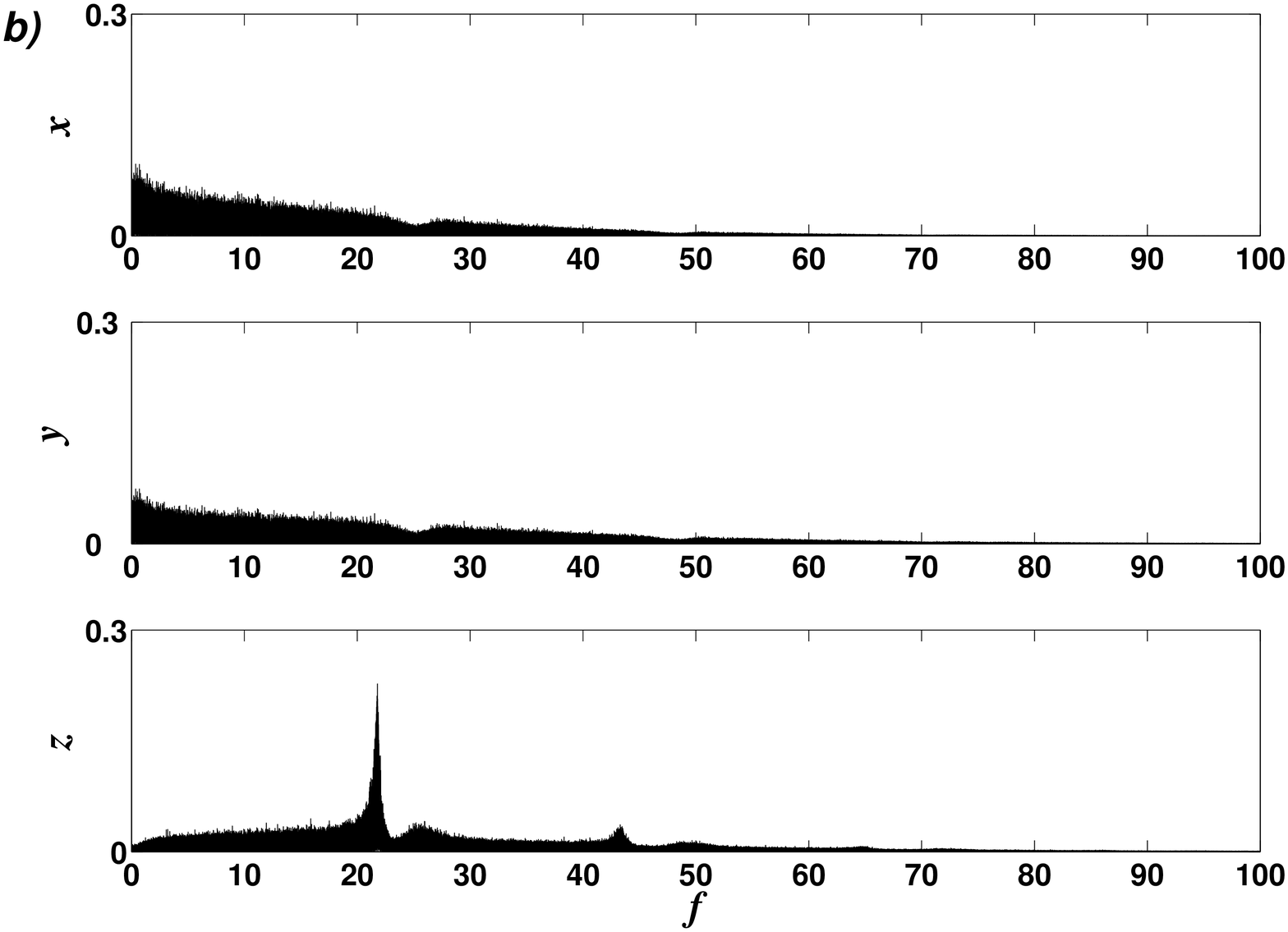}
\includegraphics[ width=0.7\textwidth]{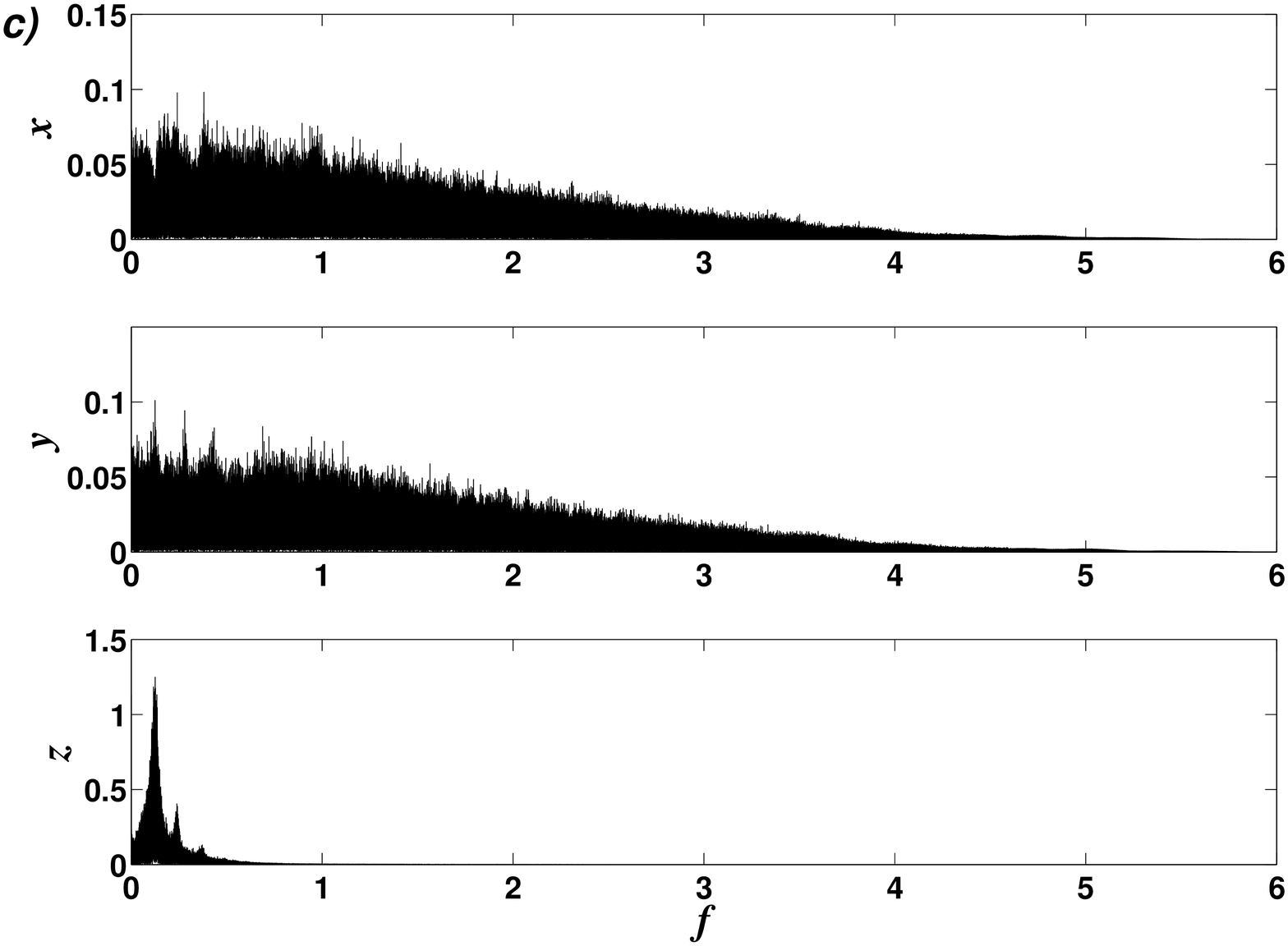}

\caption{Power spectra of the coordinates $x$, $y$, and $z$ as a function of frequency:
{\it (a)\/} Rossler System with $\tau =0.01$  with $a=0.45$, $b=2$, and $c=4$. for $\tau =0.01$, and $\Delta \tau = 0.01$.
{\it (b)\/} Lorenz System with $\sigma= 16$, $r=45.92 $, and $b=4$ for $ \tau =0.001$, and   $\Delta \tau = 0.001$.
{\it (c)\/} ${\mathbf B}_7$ System with $K~=~0.5$, $\alpha~=~7.0$, and $\epsilon~=~0.23$ for $ \tau =0.001$, and $\Delta \tau = 0.01$.
In all the cases the time series length considered had $M = 10^5~data$. }
\label{fig:PS}
\end{figure}

The first two systems exhibit strong differences that make them
interesting for this study. The Rossler-oscillator ``spends" most
of the time near the plane $y,z$ with $x\simeq0$. The time series
obtained by sampling the $x$ coordinate looks like a \textsl{delta
train} as Fig. \ref{fig:xyzt}.a illustrates. The Power Spectrum
(PS) displayed in Fig. \ref{fig:PS}.a shows that the
characteristic periodicities are very similar for all the three
coordinates, with a higher  frequency-content for $x$, produced by
its delta like shape. On the other hand, the Lorenz oscillator
displays a very different spectrum for coordinate $z$ (see Fig.
\ref{fig:PS}.b), as compared with the PS of $x$ or $y$. The
reason is that the attractor spirals from the center to the border
of one of the typical  butterfly's wings and then suddenly moves
to the other wing, as Fig. \ref{fig:xyzt}.b shows.  Both
attractors have correlation dimension $D_2$ close to $2$.
The attractor ${\mathbf B}_7$ has a  $D_2=2.719\pm0.156$
\cite{Chlouverakis2005}, covering most of the available
state-space. In Figs. \ref{fig:xyzt}.c and \ref{fig:PS}.c  the
time evolution of the three coordinates and their corresponding PS
are displayed.

The  evolution of each dynamical system was  determined by
recourse to  a variable-step Runge-Kutta-Fehlberg approach
\cite{Press1995}.
Evaluations were made using sampling periods $\tau$ ranging from
{\it  i)\/} $0.01$ to $5$, in steps  $\Delta \tau = 0.01$ for the Rossler system,
{\it ii)\/} $0.001$ to $0.3$ in steps of $\Delta
\tau = 0.001$ for the Lorenz system, and
{\it iii)\/} $0.001$ to $1$ in steps of  $\Delta \tau = 0.001$ for ${\mathbf B}_7$ system.
With these $\tau$'s one covers all the interesting regions:
(1) the over-sampled one where $\tau$ is very small as compared with
the characteristic times considered in Sec. \ref{sec:takens} and
\ref{sec:nyquist},
(2) the under-sampled region where $\tau$ is
very large as compared with the characteristic times.  For every
$\tau-$value, $10$ realizations were generated by starting from
different initial conditions. The time series for the state
variables $x$, $y$, and $z$  were stored in a matrix of $3\;
\times 10^5$, after skipping the first $10^4$ iterations in order
that transient states die out.

We employed Bandt and Pompe's procedure (with dimension $d=6$ and
time lag $T=1$) to assign a PDF to each time series. These PDF's
exhibit an entropy $H^{(BP)}$ and a complexity $C^{(BP)}$.
Functionals $H^{(hist)}$ and $C^{(hist)}$ were evaluated as well,
with a PDF obtained from the pertinent histogram. In this case,
the range covered by each variable was divided into $2^{16}$
uniformly distributed subintervals. The mean values $<C>$ and
$<H>$ were computed for all the here considered PDF-functionals
and  all the realizations. For simplicity, the symbol $<\bullet>$,
meaning mean value over realizations, is suppressed.

\subsection{Our findings}
\label{sec:findings}

The  entropy-complexity plane $H^{(BP)}\times C^{(BP)}$ is quite useful for
distinguishing between stochastic
noise and deterministic chaotic behavior was recently illustrated
in \cite{Rosso2007A}. In such a vein  we first analyze the planar
evolution  of representative points for our systems as the
sampling frequency
changes (see Fig. \ref{fig:CH}.a-c for the three sampled variables of the Rossler system,
Fig. \ref{fig:CH}.d-f for the Lorenz system, and
Fig. \ref{fig:CH}.g-i for the ${\mathbf B}_7$ system).
As 
was explained in Sec. \ref{sec:CBP}, ``limiting curves" in the plots Fig.
\ref{fig:CH} determine the allowed values of entropy and
complexity. There exists no PDF with entropy-complexity
coordinates outside the region limited by these curves
\cite{Martin2006}.

As the sample period $\tau$ decreases, the representative point
evolves in the $H^{(BP)}\times C^{(BP)}$ plane, from the right
planar region with low complexity and high entropy (under-sampling
system with small sampling frequency $f=1/\tau$) to the left
region with low complexity and low entropy (over-sampling with big
sampling frequency $f=1/\tau$). The reason for such behavior is,
as we comment in Sec. \ref{sec:takens}, that  for a very big
sampling period the dynamical correlations between consecutive
measures are lost due to the mixing behavior of the chaotic system under 
study. That is, the measures can be completely uncorrelated and
present characteristics like random behavior ($H \approx 1$ and $C
\approx 0$). On the contrary, for very small sampling periods
$\tau$ the measures will be (strongly) time-correlated so that
successive measures suggest being in the presence of  a very
ordered system ($H \approx 0$ and $C \approx 0$). The evolution of
the system in the $H^{(BP)}\times C^{(BP)}$ plane for regimes
intermediate between these two extreme sampling-period instances
will depend on the specific characteristics of the nonlinear
system under study: whether it presents mixing or not; its
chaos-degree, etc.

It is important to keep in mind that non linear correlations are
the origin of the geometric structures that constitute the
hallmark of deterministic chaos \cite{Rosso2007A}. Precisely, in
previous works $C^{(BP)}$ was seen to be a measure of the
complexity induced by these non linear correlations. Our present
results tell us that $f_M=1/t_M$ -- the sampling frequency
corresponding to the point with maximum complexity -- can be
regarded as the \textsl{optimum sampling frequency}, i.e, the
minimum sampling frequency that retains all the information
concerning these structures. Higher sampling frequencies imply an
\textit{oversampling}, producing unnecessary long files to cover
the full attractor, while lower sampling frequencies make the time
series to lose some vital information concerning  nonlinear
correlations. In the oversampled case, specific  $d$-length
ordinal patterns $(0,1,\cdots,d-1)$ and $(d-1,\cdots,1,0)$  appear
more frequently than any other ordinal pattern, causing a low
entropy $H^{(BP)}$ and a low complexity $C^{(BP)}$.
Points  in the region with high entropy and low complexity of
the $H^{(BP)}\times C^{(BP)}$ plane correspond to
\textsl{undersampled trajectories}  of the continuous system.
The samples are not correlated at all and the  system behaves randomly.
Consequently, all the ordinal patterns appear with almost the same frequency,
as the ensuing high entropies $H^{(BP)}$ and  low complexities $C^{(BP)}$ reveal.
The evolution of the representative point in the plane $H^{(BP)}\times C^{(BP)}$
near the top limiting curve is typical of chaotic systems, as demonstrated by the authors
of \cite{Rosso2007A}.

Let us stress the importance of using the Bandt and Pompe  procedure to obtain the all-important PDF.
A well-known result is that any chaotic map shares with its iterated maps the same histogram
\cite{Setti2002}.
This property is the basis of the \textsl{skipping randomizing technique} \cite{DeMicco2008,Callegari2003A}.
For that reason De Micco {\it et al.} \cite{DeMicco2008} proposed the use of two different PDFs,
as relevant for testing
{\it (i)\/} the uniformity of $\mu(x)$ (the invariant measure) and
{\it (ii)\/} the mixing constant $r_{mix}$ pertaining to a chaotic time series, as a means to generate
pseudo random numbers.
They proceeded to employ
{\it (a)\/} a PDF $P_1$ based on time series' histograms and
{\it (b)\/} another PDF $P_2$ based on ordinal patterns
(permutation ordering) that derives from appealing to the already cited Bandt and Pompe method \cite{Pompe2002}.
One may conjecture that continuous systems exhibit an analogous property, viz., the
sampling process does not significantly change the histogram of the time series and, consequently,
any quantifier based on the PDF histogram is not sensitive to the sampling frequency.

Such conjecture  is validated  in Fig. \ref{fig:BPvsHist}, where
the statistical complexities $C^{(BP)}$ and $C^{(hist)}$  are
depicted as functions of the sampling frequency for
the $x$-coordinates of the three systems (see Fig. \ref{fig:BPvsHist}.a
for Rossler's instance, Fig. \ref{fig:BPvsHist}.b for Lorenz'
case, and Fig. \ref{fig:BPvsHist}.c for ${\mathbf B}_7$). A similar result
is obtained if $C^{(BP)}$ and $C^{(hist)}$ are evaluated for the
other coordinates $y$ and $z$, and also if mean values
$C^{(BP)}_{(xyz)} =[ C^{(BP)}_{(x)}+C^{(BP)}_{(y)}+C^{(BP)}_{(z)}
]/3$ and $C^{(hist)}_{(xyz)} =[
C^{(hist)}_{(x)}+C^{(hist)}_{(y)}+C^{(hist)}_{(z)} ]/3$ are used.

Consequently, we conclude that the
usual  histogram technique is not
useful to analyze the effects of the sampling frequency and that a
causal PDF (here the Bandt and Pompe one) \emph{must} be employed
instead. Note that in Figs. \ref{fig:CH} and \ref{fig:BPvsHist}
many irregularities of the trajectories become visible. These are
produced by the nontrivial relation between $C^{(BP)}$ and
$H^{(BP)}$, and diminish as the number of realizations increases.
Note that in Fig. \ref{fig:BPvsHist}, where $C^{(BP)}$ is
represented as a function of $f= 1/\tau$, a maximum is easily detected.

Table \ref{tab:chartimes2} summarizes our results. $t_M^{(x)}$,
$t_M^{(y)}$, and $t_M^{(z)}$  are the inverses of the \textsl{
sampling frequency}  for each coordinate producing the maximum
$C^{(BP)}$. We have evaluated $t_M^{(x)}$, $t_M^{(y)}$, and
$t_M^{(z)}$  for each realization and then  determined the
associated mean values and standard deviations shown in columns
$1$ to $6$.  The mean value of $C^{(BP)}$ over coordinates $x$,
$y$, and $z$ also exhibits  a maximum for the $t_M$ indicated in
column $7$ ($<x, y, z>$).  All the characteristic times described
in Sec. \ref{sec:takens} and \ref{sec:nyquist}  are displayed in
Table \ref{tab:chartimes2}. Column $8$ displays  the ratios
between each characteristic time and the $t_M$ of column $7$.
Figures closer to unity in column $8$ (within each class of
characteristic times) are emphasized by using bold-types.

As we mention in Sec. \ref{sec:CBP}, the  normalized Shannon
entropy and statistical complexity evaluated with PDF-Bandt and
Pompe are not  system invariant (neither are other quantifiers
derived by Information Theory). The values of these quantifiers
depend on how the PDF is evaluated. However, they provide very
useful information about the dynamics of the system under study.
It is clear then that $H^{(BP)}$ and $C^{(BP)}$  and the
quantities derived from them could attain different values
according  with the specific time series considered (i.e. time
series generated by the different coordinates of the system or a
combination of them) even if it is reasonable to expect that their
corresponding values  do not exhibit a strong dispersion. This
assumption is clearly  confirmed if we look at the  values
obtained for $t_M$ in Table \ref{tab:chartimes2}, evaluated for
different times series. As expected, the same kind of behavior is
obtained in the case of $t_I$ (Average Mutual Information).

From inspection of the results displayed in Tab. \ref{tab:chartimes2} we
see that the time $t_M$ yields similar values to those obtained from
the evaluation of the Average Mutual Information $t_I$. The
largest  difference between $t_M$ and $t_I$ (see  column 8 in Tab.
\ref{tab:chartimes2}) is observed for the ${\mathbf B}_7$ system.
Such disparity can be attributed to the fact that this system
possesses an attractor with $D_2 \approx 3$ (that covers most of
the state-space) in contrast with the other two systems considered
here, the Lorenz and Rossler ones, which have $D_2 \approx 2$
\cite{Kantz1999,Abarbanel1996}.

With regards to  $t_M$ and the others special
characteristic times (for our three dynamical systems) we can state that,
in the Rossler-case, such characteristic times are,
respectively, the first minimum of the mutual information $t_3$,
the first zero of the autocorrelation $t_0$, and $t_{NS}$, for
$\alpha=0.85$. In the the Lorenz-instance the characteristic times
closest to $t_M$ are: the first minimum of the mutual information
$t_3$, the first change of curvature of the autocorrelation $t_2$,
and $t_{NS}$ for $\alpha=0.9$. Finally, for  ${\mathbf B}_7$ the
characteristic times closest to $t_M$ are: $t_{1/e}$, and $t_{NS}$
for $\alpha=0.85$. The first minimum of the mutual information
appears far from the $t_M-$location. The shape of $I_m$ (discrete
mutual information) is very irregular, as shown in Fig.
\ref{fig:IM}.c, which explains why
such features takes place.

For all the systems, the Nyquist criterion provides 
smaller characteristic times if $\alpha\ge0.95$ is used instead,
illustrating thus on  the difference between the \textsl{exact
reconstruction} that is the goal of the Nyquist-Shannon Theorem
and the \textsl{approximate attractor reconstruction} that relies
on Takens' theorem. Note that the autocorrelation function is
based just on linear statistics and does not contain any
information concerning nonlinear dynamical correlations. Kantz and Schreiber
\cite{Kantz1999} stressed the drawback of using  linear statistics for
the Lorenz system, with a typical trajectory staying some time on
one wing of the butterfly attractor and spiralling then from the
inside to the border before jumping to the other wing. The
autocorrelation does not reveal the importance of the period of
such ``jumping" between wings, but the mutual information does so.
Also, the autocorrelation of the squared signal is able to detect
this periodic change \cite{Kantz1999}.

To assess  the quality of the Takens' reconstruction process using
$t_M$,  $D_2$ is evaluated for an embedding dimension ranging from
$d=1$ to $10$ (see Fig. \ref{fig:d2A9}).  The pertinent estimation of
$D_2$ is compared with values reported in the literature for the
continuous system \cite{Kantz1999,Sprott2005,Sprott2001}. The
subroutine \textit{d2} of the TISEAN package \cite{Hegger1999} was used
for calculating $D_2$.

Let us stress that the estimation of $D_2$ poses a very delicate
 problem, so that we refer the readers to the documentation
available for the TISEAN package and also to the excellent book by
Kantz and Schreiber \cite{Kantz1999}. Figs. \ref{fig:d2A9} show
that $\tau=t_M$  produces a time series with a $D_2$ that turns
out to be  a better estimation for the Correlation Dimension of
the continuous system than those obtained with higher
(under-sampled) or lower (over-sampled) values for $\tau$. Similar
results were obtained when we consider the evaluation of the
maximum Lyapunov exponent.

It is interesting to remark that the time $t_M$,  for which one
detects a maximum in the entropy-complexity plane $H^{(BP)} \times
C^{(BP)}$, is unique for the three dynamical systems analyzed in
the present work.
We associate this sample-time with the 
one needed  to capture all the relevant information related to the
dynamics'  nonlinear correlations.
Recently Soriano {\it et al.\/} \cite{Soriano2011}  theoretically
and experimentally studied a chaotic semiconductor laser with
optical feedback. This is an extremely interesting system because
dynamical systems with time delay, like the Mackey-Glass or a
laser with delay feedback, exhibit different relevant time scales
\cite{Zunino2010,Soriano2011}. Consequently, additional
complexity-maxima could be expected (in addition to the one
encountered in dealing with the chaotic systems of this paper). In
the case of the laser studied in \cite{Soriano2011} the main time
scales are: i) $\tau^*_S$, the feedback-time providing the largest
time scale,  ii) $T_{RO}$, that is the relaxation oscillation
providing a shorter time scale, and finally iii)  chaotic
oscillations, that  govern the fastest time scale.
Soriano {\it et al.\/} experimentally confirmed that our
prescription for $t_M$ corresponds to a maximum of $C^{(BP)}$, as
found in all the chaotic systems studied in this paper (see Fig. 10
in \cite{Soriano2011}). To our knowledge, this is the first
controlled experiment where our prescription has been to hold.

Sharing the point of view of Abarbanel \cite{Abarbanel1996}, our
prescription for the sample period $\tau=t_M$ is based on the
consideration of a fundamental aspect of chaos, namely, the
generation of information. Thus, our choice is made by taking into
account an important property of the system we wish to describe.
Remark that stable linear systems generate zero information. In
consequence,  information-generation \cite{Abarbanel1996} is a
property of nonlinear dynamics not shared by linear evolutions.

Finally, another issue to be aware of is the influence of noise.
To verify the robustness of the  criterion here advanced, gaussian
noise was added to each system and the value of $t_M$ was obtained
as a
function of the signal to noise $S/N$ ratio ($\sigma$) of this
gaussian noise. %
In all cases, the value of $t_M$ remained constant
(up to three significant decimal digits) for $S/N >20~dB$. If the
noise-strength is higher, $t_M$ suddenly decreases as the
representative point in the plane $H^{(BP)} \times C^{(BP)}$ moves
to the rightwards region (stochastic region), where
ordering-patterns are strongly affected by noise.

\section{Conclusions}
\label{sec:Conclusions}

Our main goal here  was to show that a particular version of the
Statistical Complexity Measure $C^{(BP)}$,  evaluated by recourse
to the probability distribution obtained via the Bandt and Pompe
procedure, allows one to determine, under the light of a Takens'
reconstruction procedure,
convenient sampling periods.
This is done so as to get  time series useful for investigating  chaotic  behavior.

This sampling-period was called $t_M$,
and the procedure was 
illustrated  via a detailed consideration of three paradigmatic
systems.
On the basis of these significant examples  we conjecture that our
optimality criterion may be of general application for chaotic
systems, since $C^{(BP)}$ is a measure of the geometric structures
produced by nonlinear correlations,  always present in this class
of dynamical systems.
We showed that $t_M$ is compatible with those specific times recommended in the literature as
adequate delay-ones in Takens' reconstruction.
Our results closely approach the exact Nyquist-Shannon
reconstruction. This is so because  high frequencies of the
Fourier Spectrum are due to chaotic oscillations, and this is the
case for  the systems studied here.
 To our knowledge, controlled experiments for different sampling times of the measured variables
and large numbers of data
 have not yet been reported in the literature and the highest sampling frequency
allowed by the digital acquisition system is the one usually
adopted (as an exception see  \cite{Soriano2011}). 
In the light of our present results  it would be better, in practice, to assess the correct $t_M-$value
for a clever choosing of the sampling period. This permits one to cover the whole attractor-basin
and retain in the reconstruction process the most  important properties
of  chaotic systems. %

In view of the extensive use of digital acquisition systems in all kinds of
experiments, we hope that experimentalists may consider  the
present  contribution as a practical tool in their activities and
may thus be able to validate our proposal.

\section*{Acknowledgments}
This work was partially  supported  by  the  Consejo  Nacional de
Investigaciones Cient\'{\i}ficas y T\'ecnicas (CONICET), Argentina
(PIP 112-200801-01420), ANPCyT and UNMDP  Argentina (PICT 11-21409/04).
O. A. Rosso gratefully acknowledge support from CAPES, PVE fellowship, Brazil.

\begin{figure}
\newpage
\includegraphics[ width=0.7\textwidth]{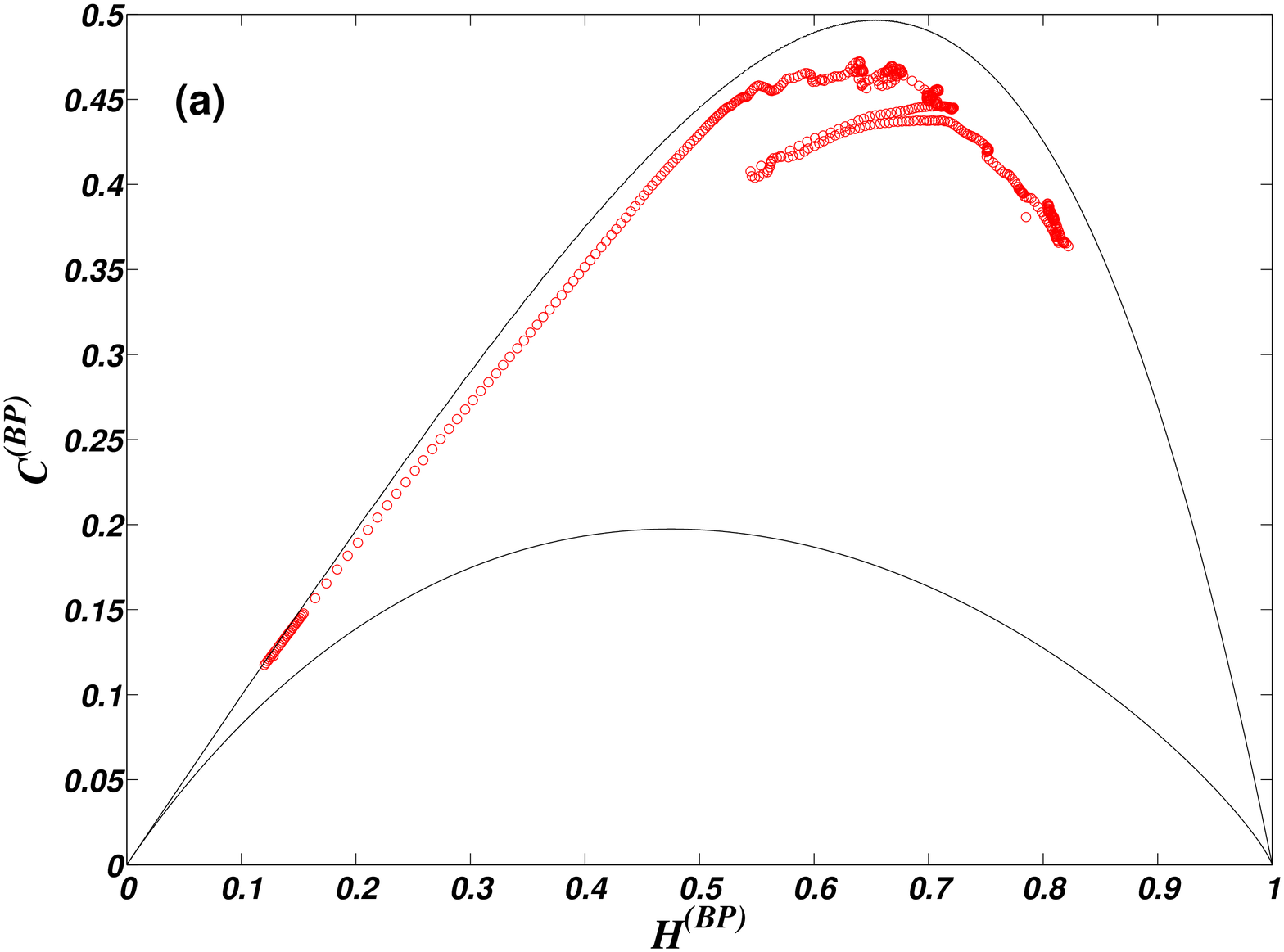}
\includegraphics[ width=0.7\textwidth]{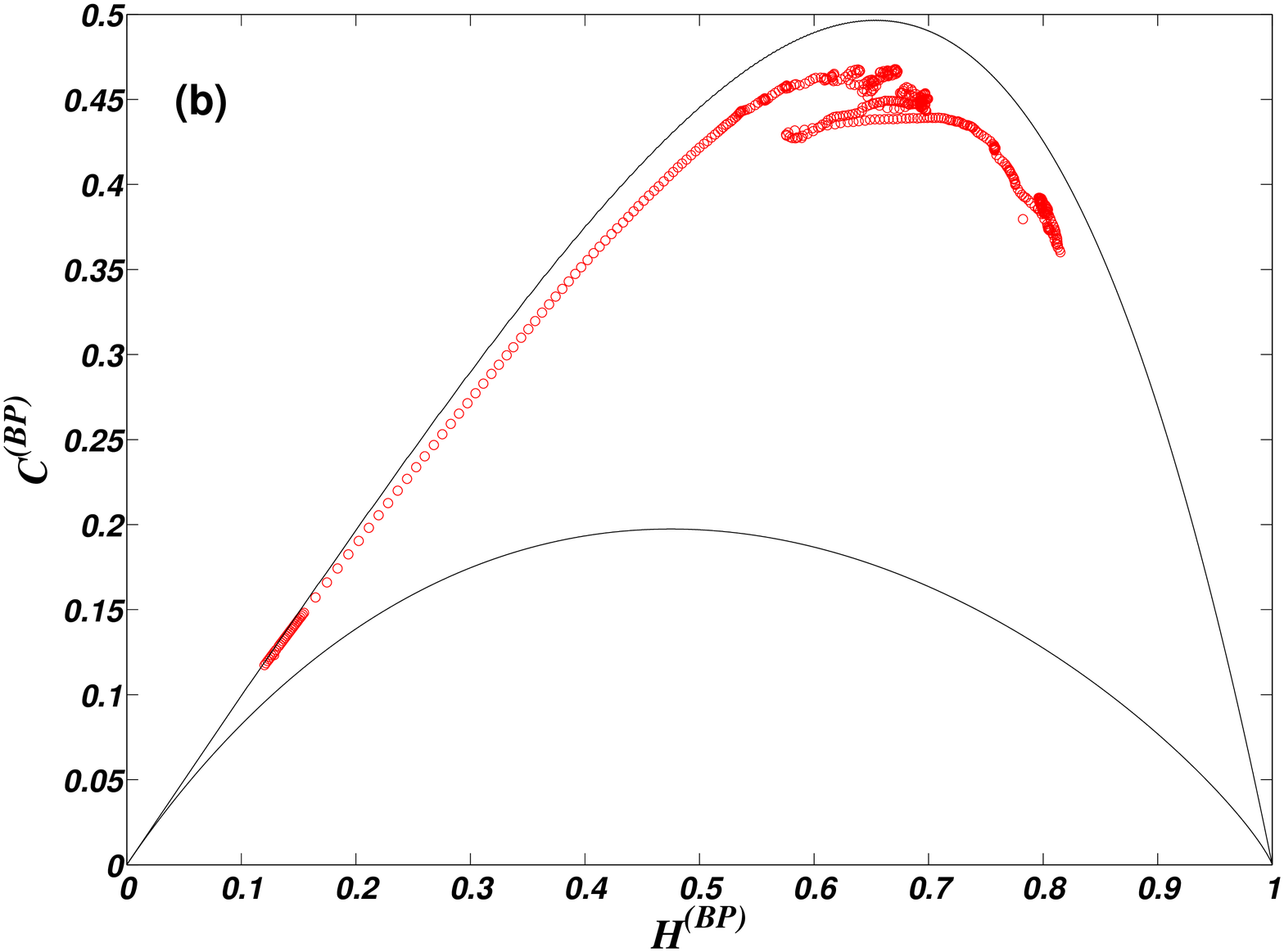}
\includegraphics[ width=0.7\textwidth]{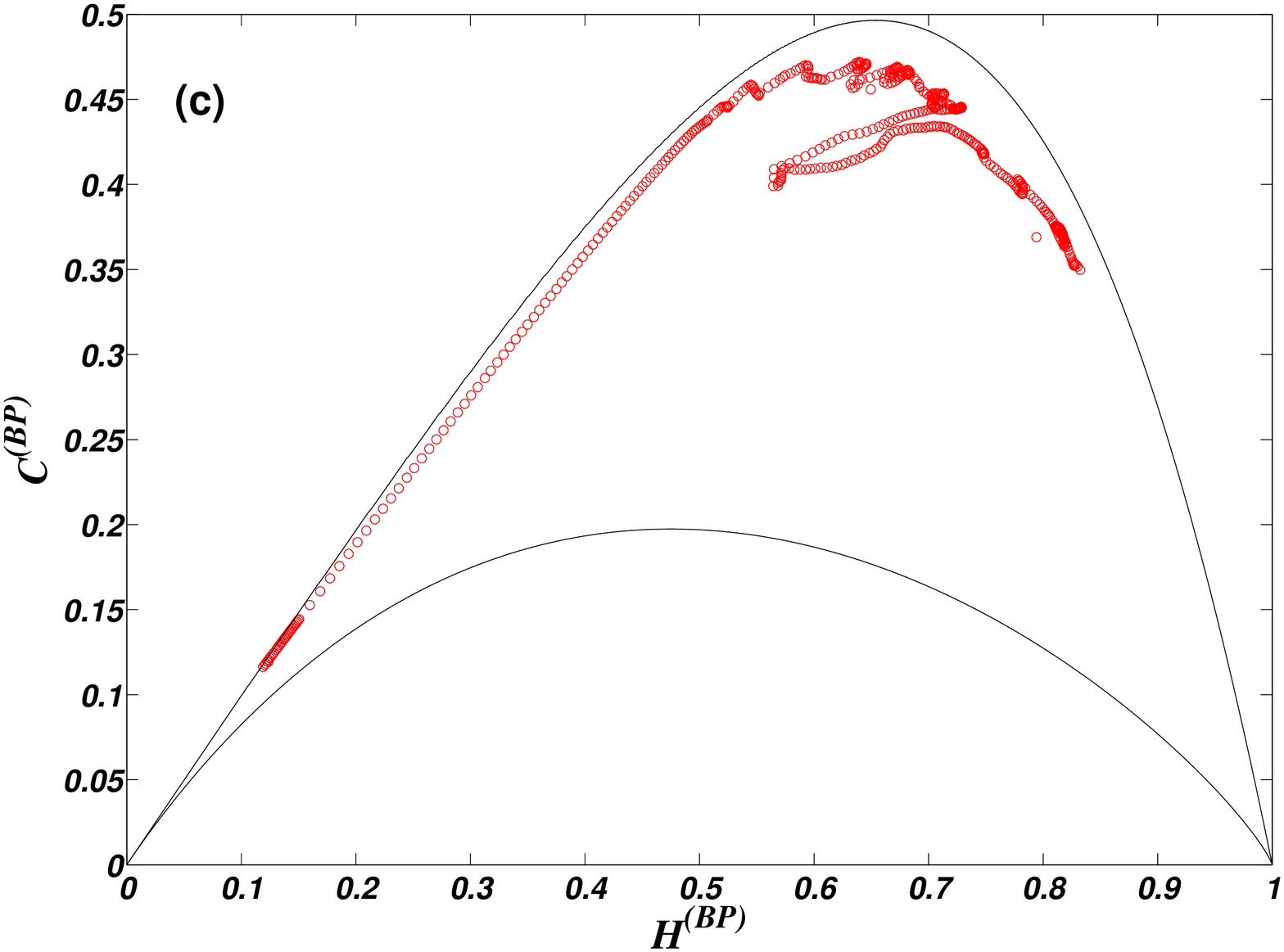}

\caption{
Representation on the entropy-complexity plane:
{\it (a-c)\/} Rossler variables $x$, $y$, and $z$ for $0.01 \leq \tau\leq 5$, $\Delta \tau = 0.01$,
with  $a=0.45$, $b=2$,  $c=4$. 
{\it (d-f)\/} Lorenz variables  $x$, $y$, and $z$ for  $0.001 \leq
\tau\leq 0.3$, $\Delta \tau = 0.001$, with $\sigma= 16$, $r=45.92$, and $b=4$.
{\it (g-i)\/} ${\mathbf B}_7$ variables $x$, $y$, and $z$ for $0.001 \leq \tau\leq 3$, $\Delta \tau = 0.001$, with
$K~=~0.5$, $\alpha~=~7.0$, and $\epsilon~=~0.23$.
Undersampling corresponds to the high $H^{(BP)}$ and low $C^{(BP)}$ region, and
oversampling to low  $H^{(BP)}$  and low $C^{(BP)}$. In all the
cases the time series length  considered had $M=10^5~data$. The
continuous lines represent, respectively, the maximum and minimum
complexity values (for $d=6$) for a fixed value of the entropy. }
\label{fig:CH}
\end{figure}

\setcounter{figure}{2}
\begin{figure}
\newpage
\includegraphics[ width=0.7\textwidth]{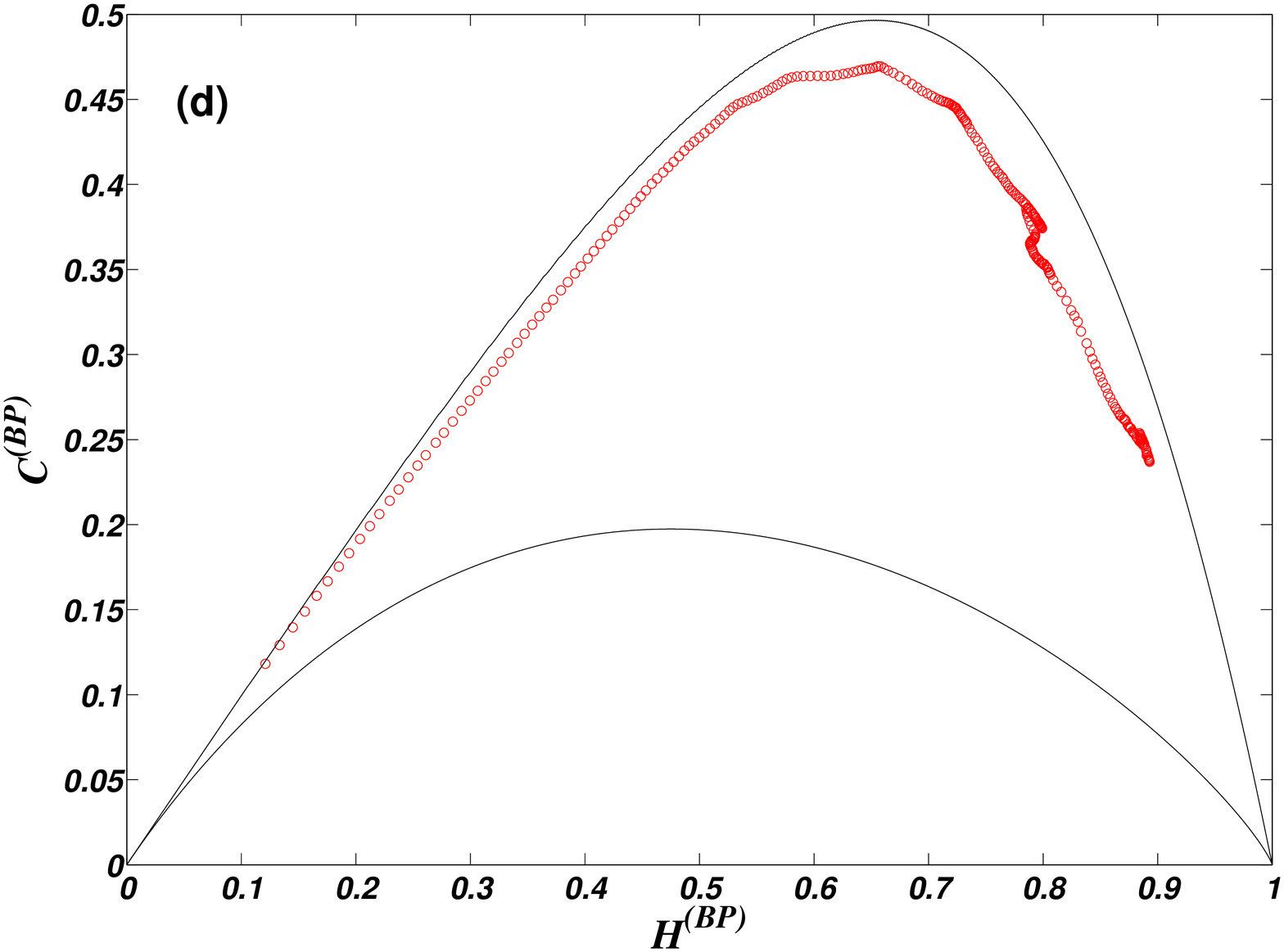}
\includegraphics[ width=0.7\textwidth]{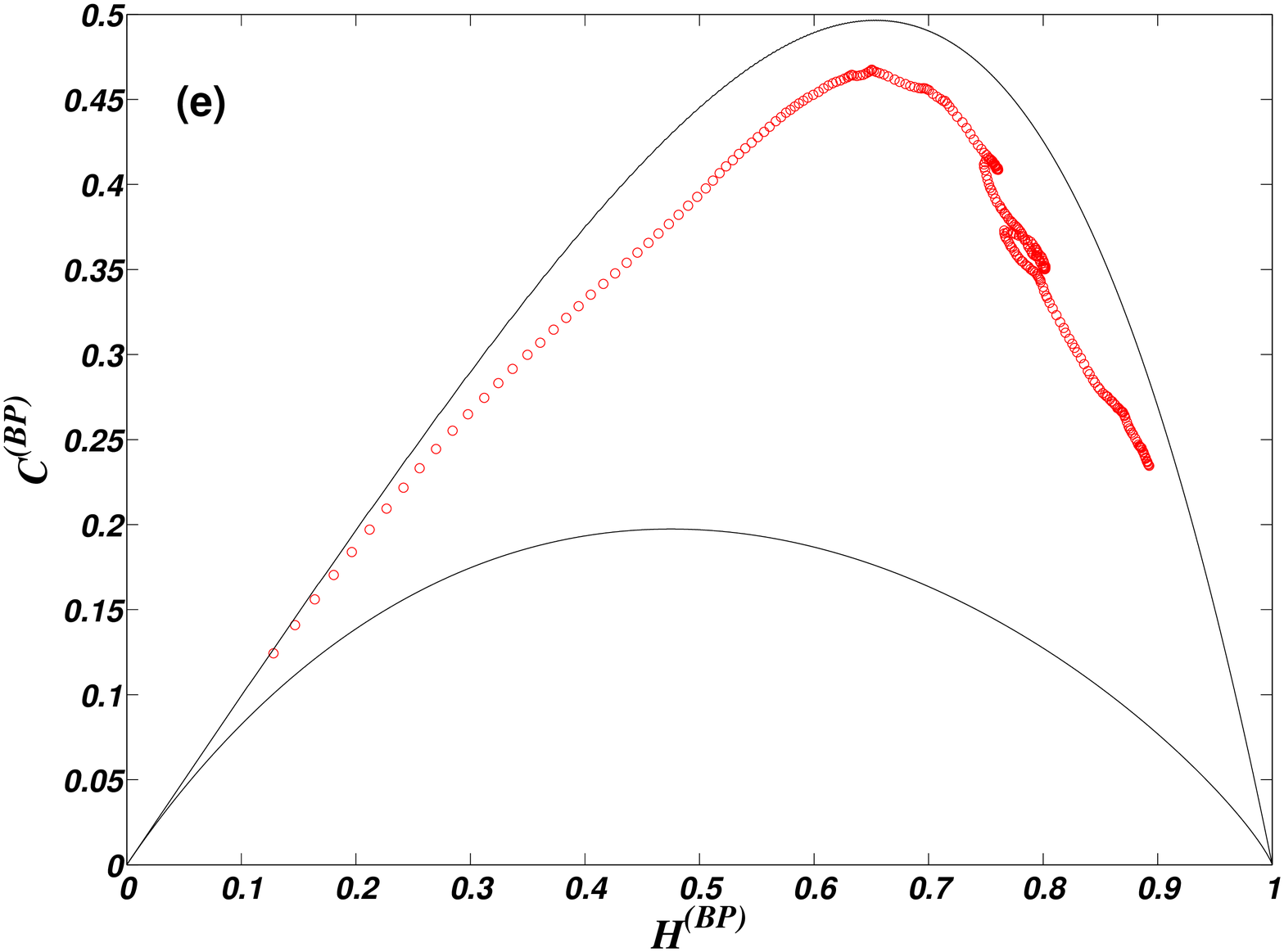}
\includegraphics[ width=0.7\textwidth]{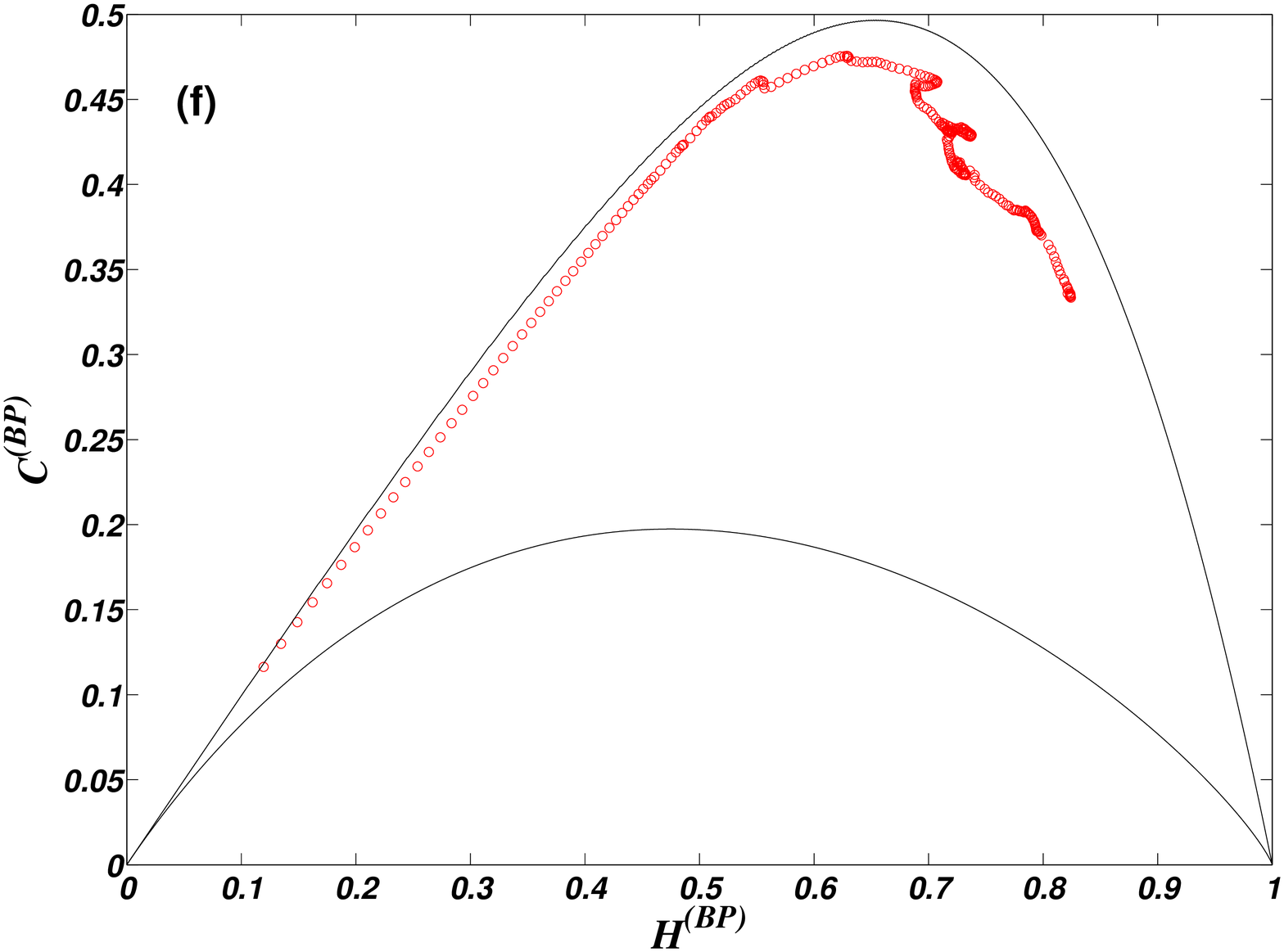}
\caption{Continuation}
\label{fig:CH1}
\end{figure}

\setcounter{figure}{2}
\begin{figure}
\newpage
\includegraphics[ width=0.7\textwidth]{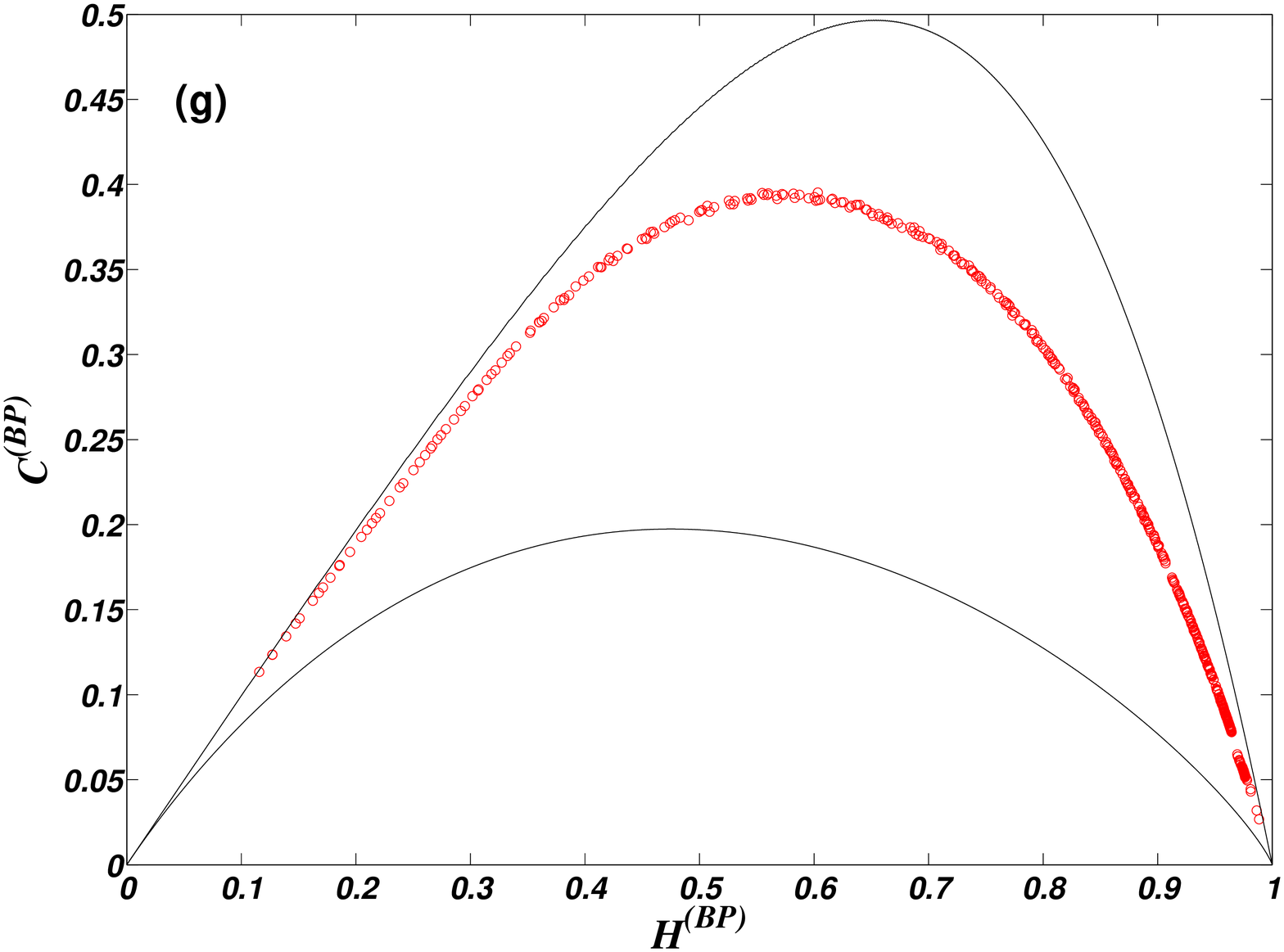}
\includegraphics[ width=0.7\textwidth]{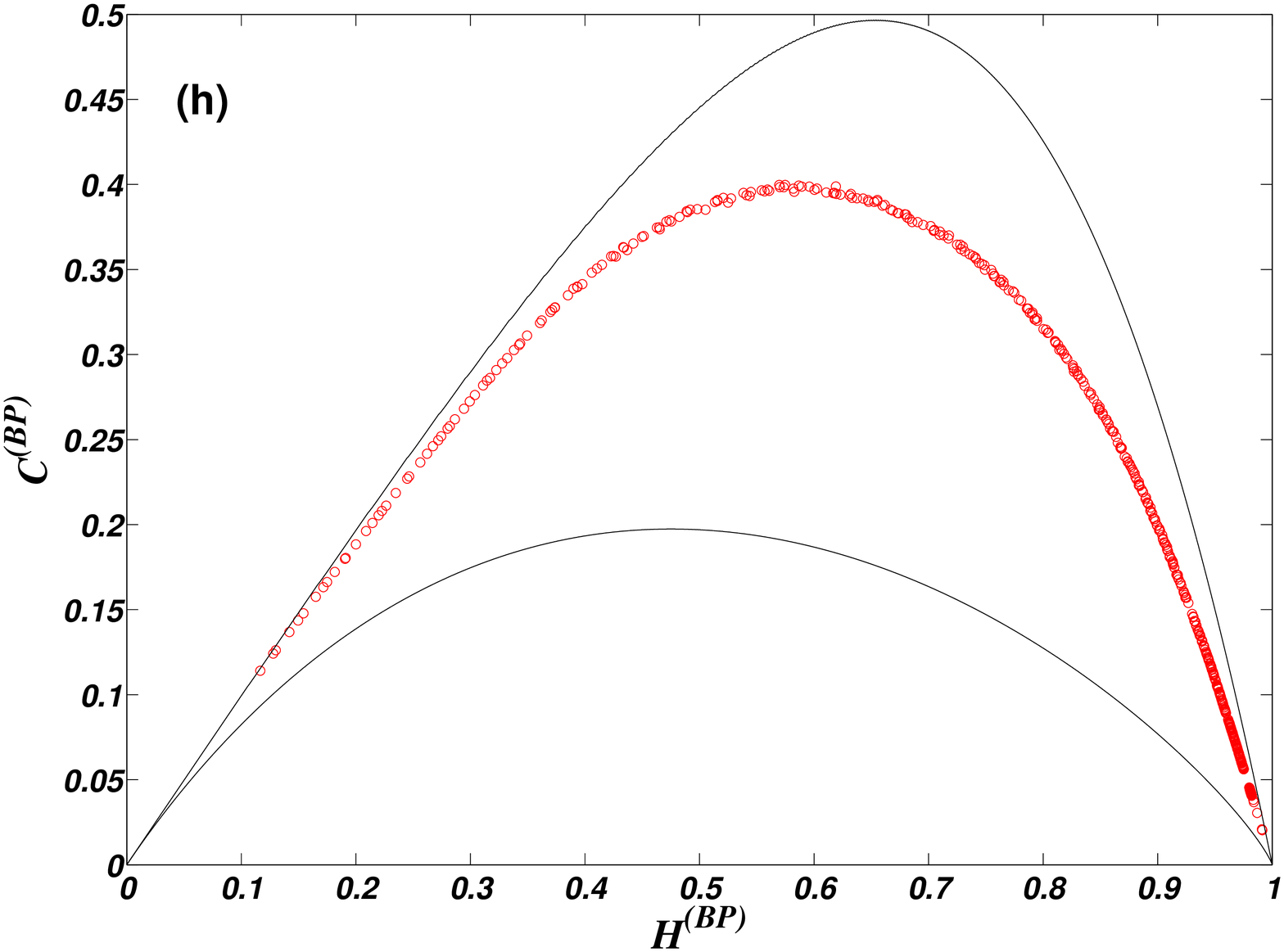}
\includegraphics[ width=0.7\textwidth]{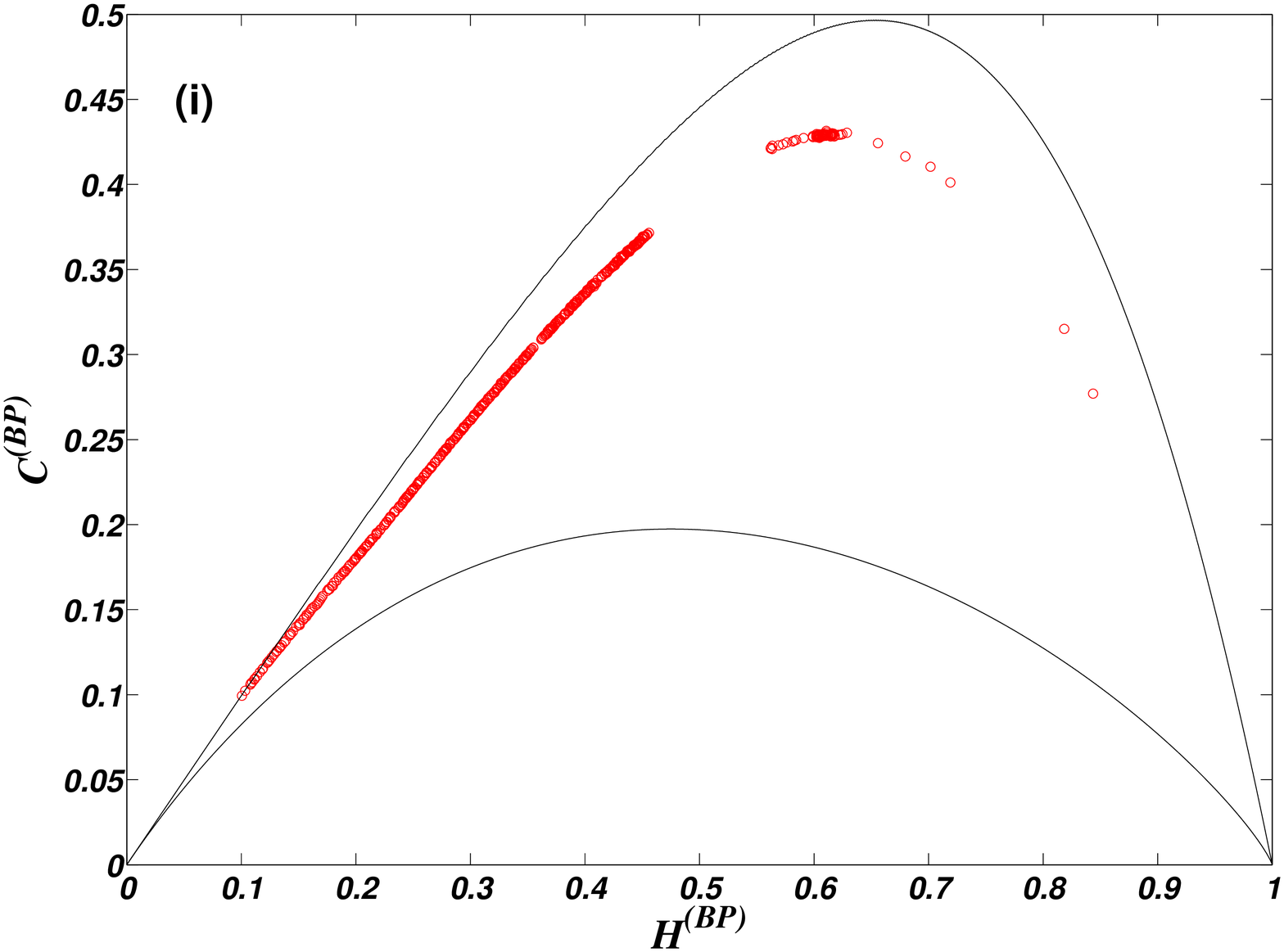}
\caption{Continuation}
\label{fig:CH2}
\end{figure}

\begin{figure}
\newpage
\includegraphics[ width=0.7\textwidth]{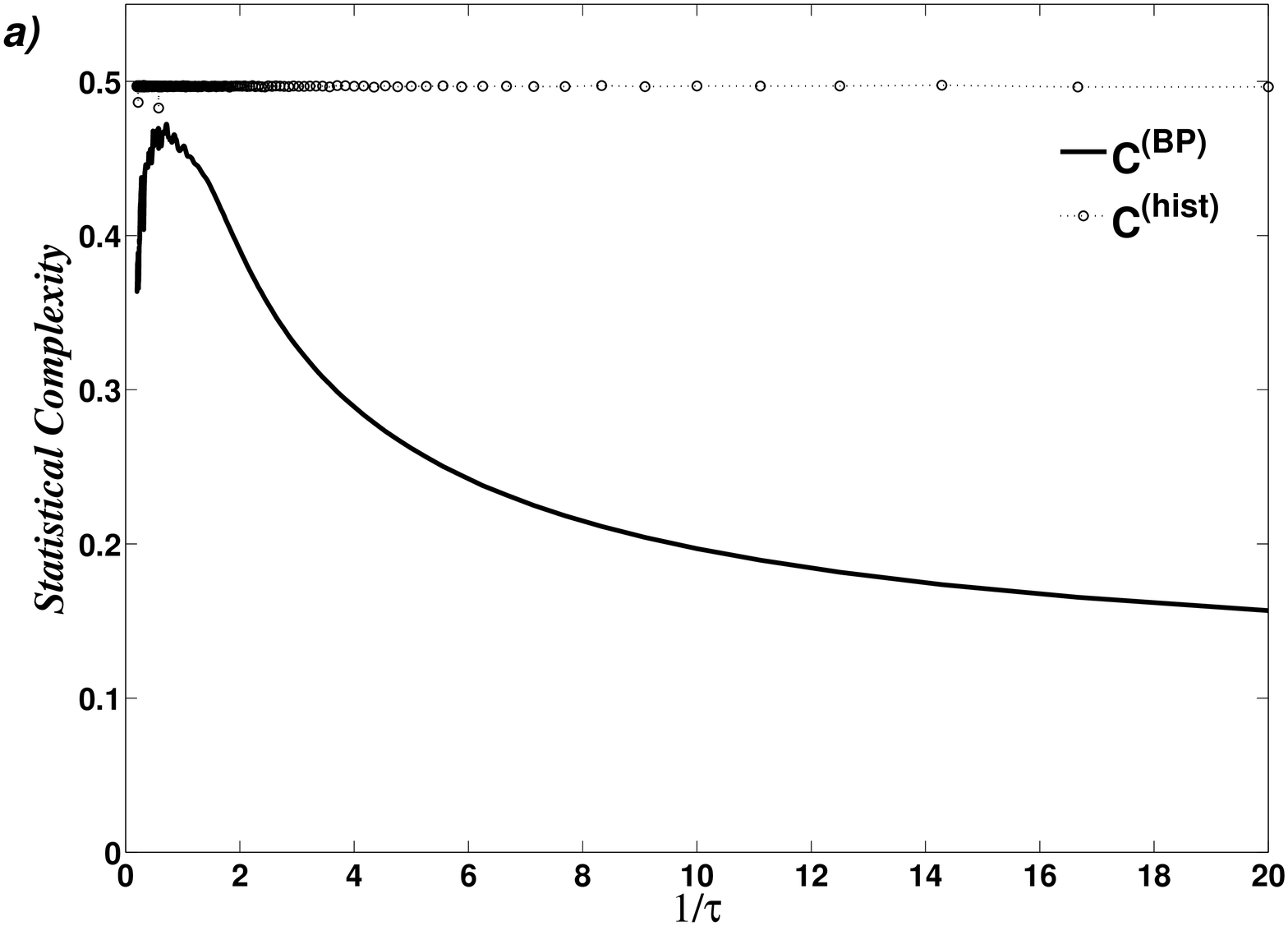}
\includegraphics[ width=0.7\textwidth]{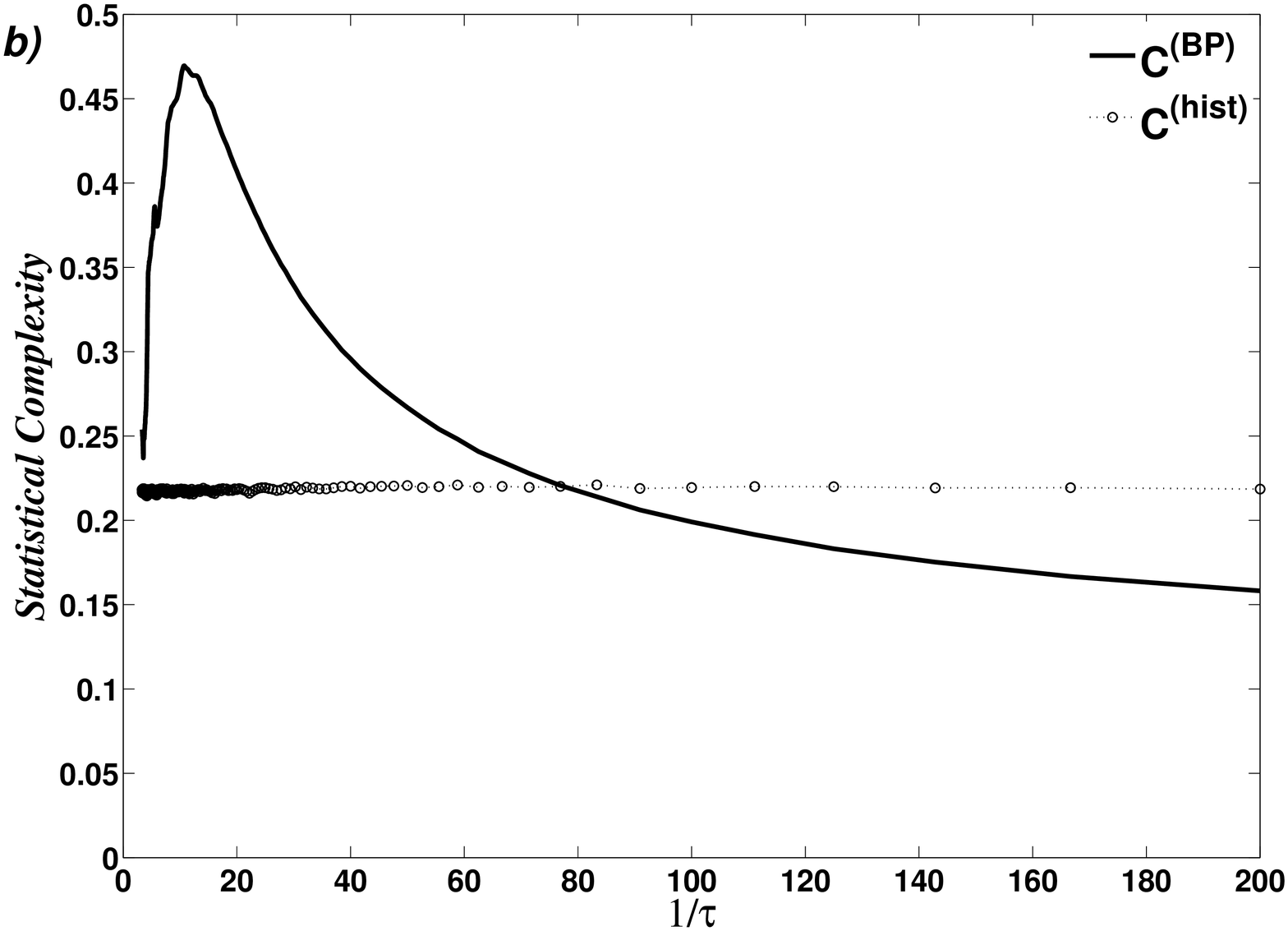}
\includegraphics[ width=0.7\textwidth]{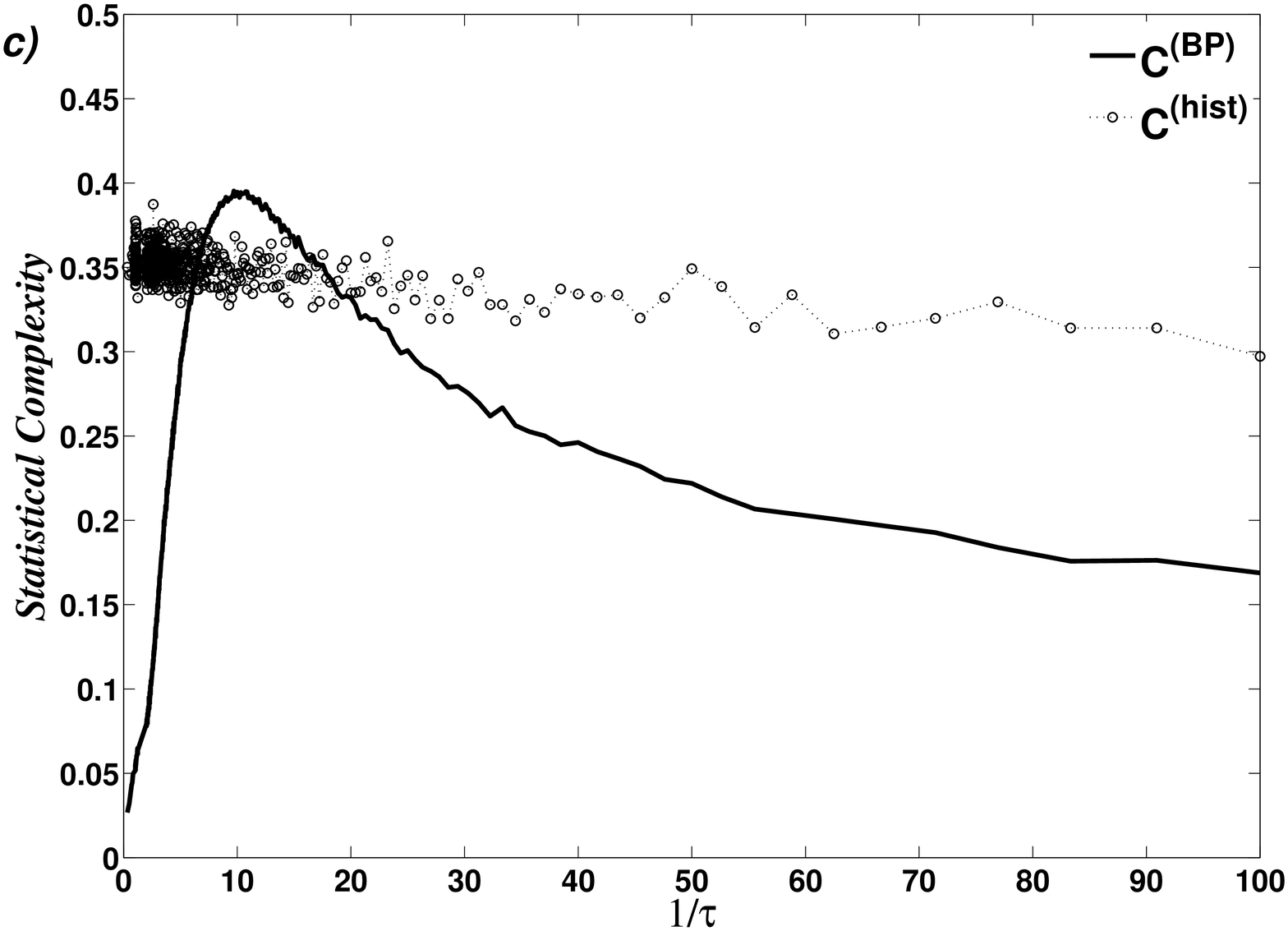}

\caption{ Intensive statistical complexity measures $C^{(BP)}$ and
$C^{(hist)}$ for the coordinate $x$ of the three systems.
{\it (a)\/} Rossler system,
{\it (b)\/} Lorenz system and
{\it (c)\/} ${\mathbf B}_7$ system one.
$C^{(BP)}$ is evaluated by using the
PDF obtained with the Bandt and Pompe prescription with $d=6$.
$C^{(hist)}$  is evaluated by using the PDF obtained from the
histogram of $x$, using $2^{16}$ bins. $C^{(hist)}$  is almost
constant for all values of $\tau$. In all the cases the time
series length  considered had $M = 10^5~data$. The use of the BP
prescription is of the essence in getting the complexity-maxima.
The reason is that a change in the sampling frequency does not
change the invariant measure of each variable of the chaotic
attractor. The situation is identical to that of the skipping
procedure in pseudo random number generators based on chaotic maps
\cite{DeMicco2008}. }
\label{fig:BPvsHist}
\end{figure}

\begin{figure}
\newpage
\includegraphics[ width=0.8\textwidth]{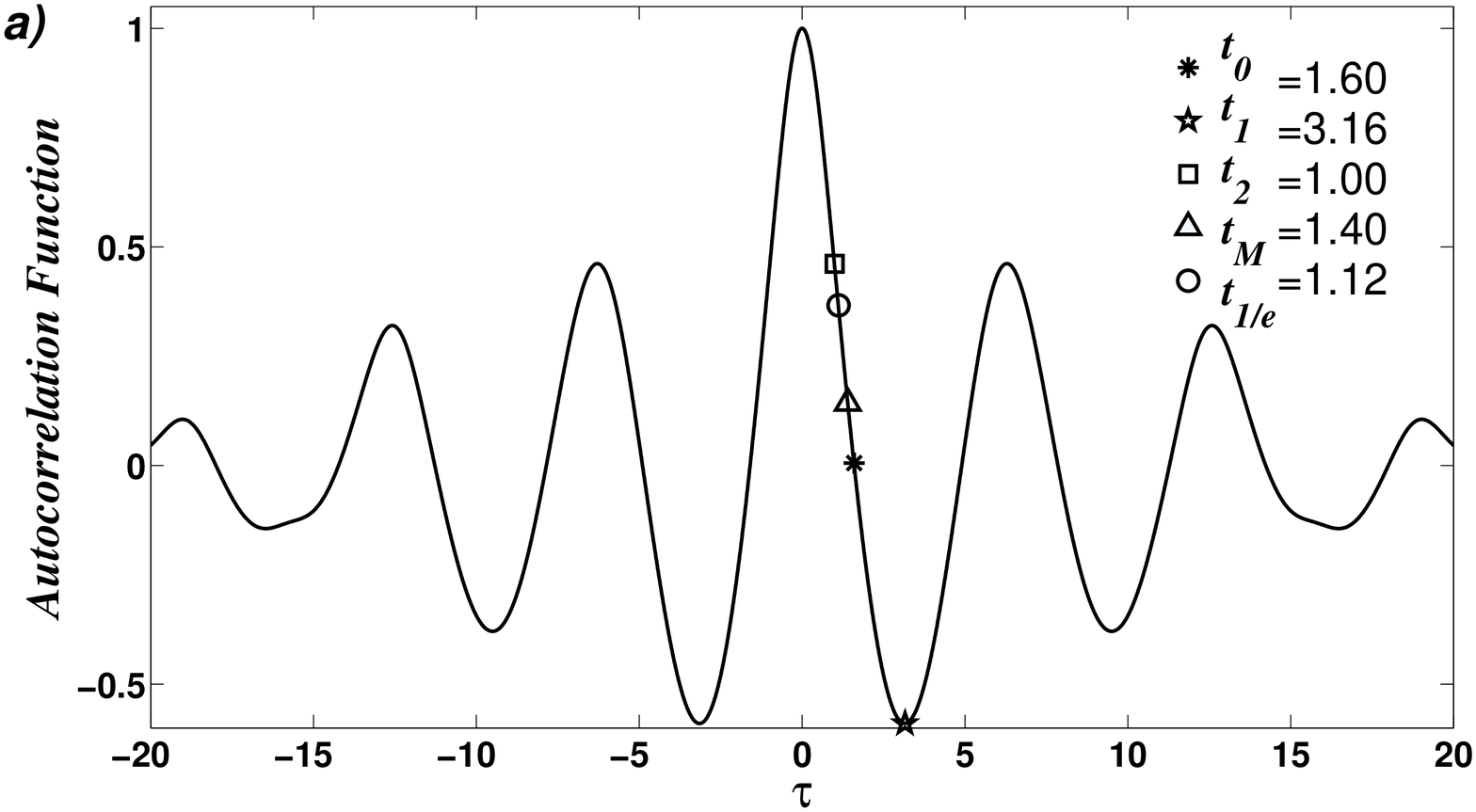}
\includegraphics[ width=0.8\textwidth]{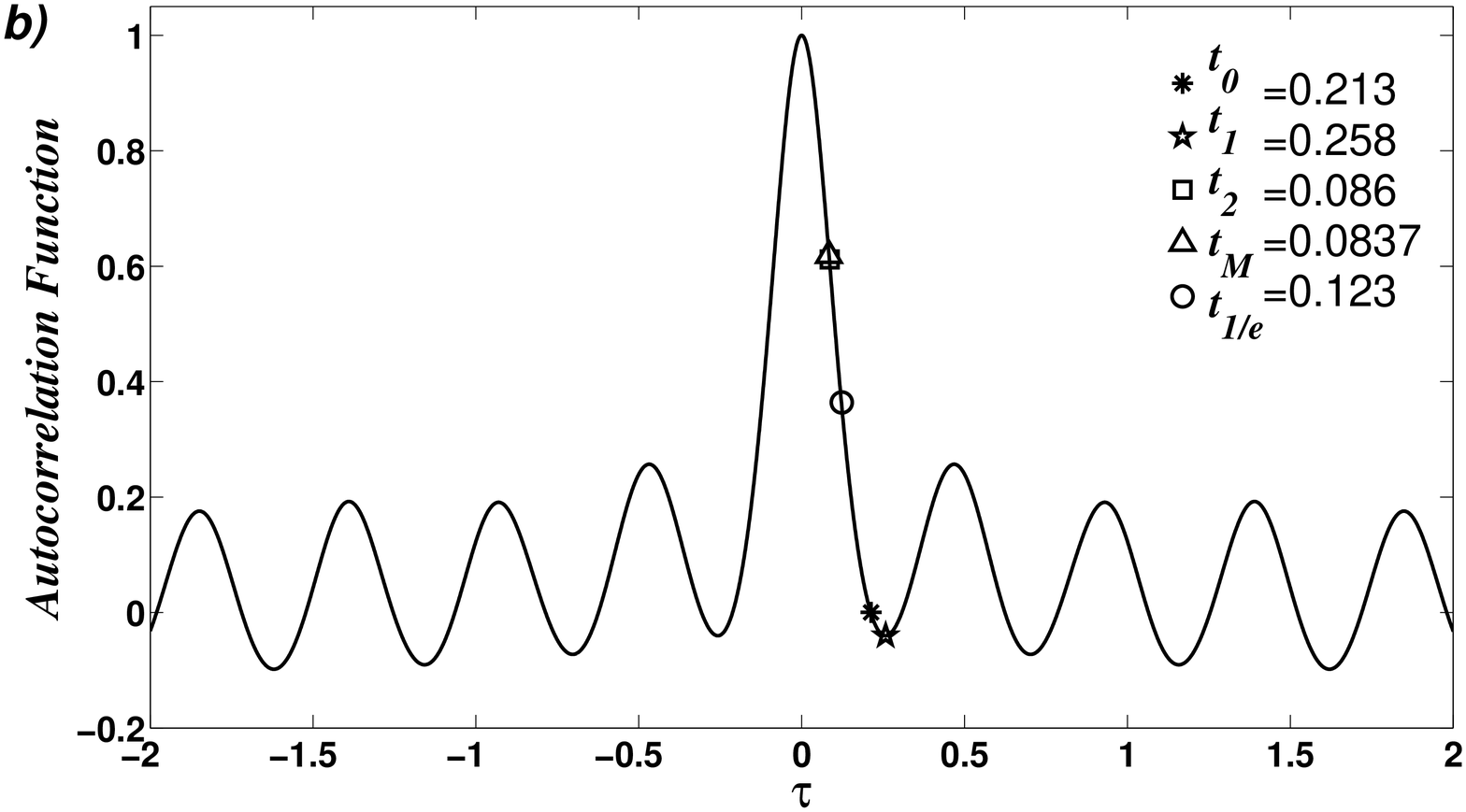}
\includegraphics[ width=0.8\textwidth]{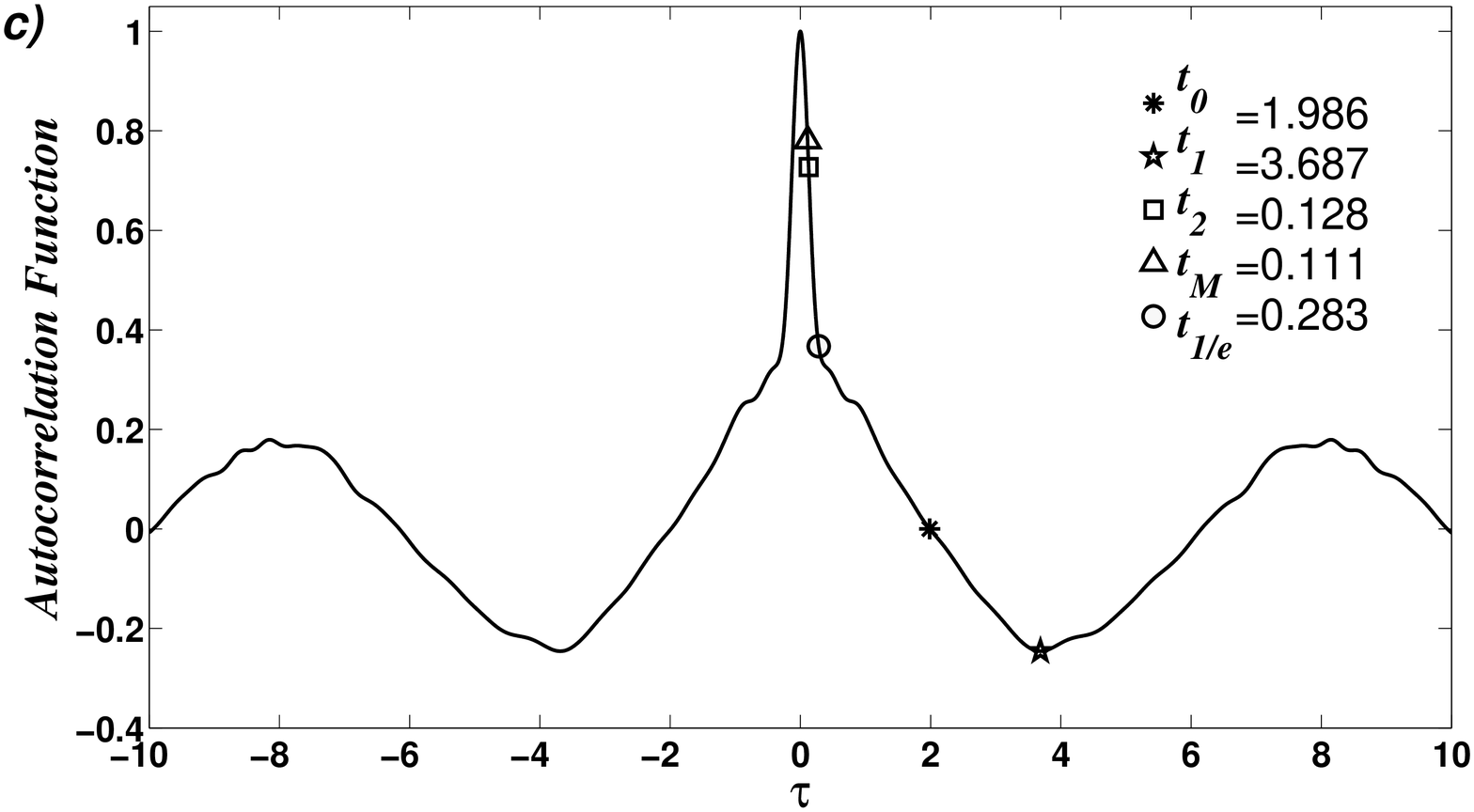}

\caption{ Discrete autocorrelation function $R_m$ for:
{\it (a)\/} Rossler System with  $a=0.45$, $b=2$,  $c=4$.
{\it (b)\/} Lorenz System with $\sigma= 16$, $r=45.92 $, and $b=4$.
{\it (c)\/} ${\mathbf B}_7$ System with $K~=~0.5$, $\alpha~=~7.0$, and $\epsilon~=~0.23$.
In all the cases the time series length
considered had $M = 10^5~data$. We show both the characteristic
time induced by $C^{(BP)}$ ($d=6$), that is $t_M$, and its most
approximate characteristic time induced by $R_m$. }
\label{fig:AC}
\end{figure}

\begin{figure}
\newpage
\includegraphics[ width=0.8\textwidth]{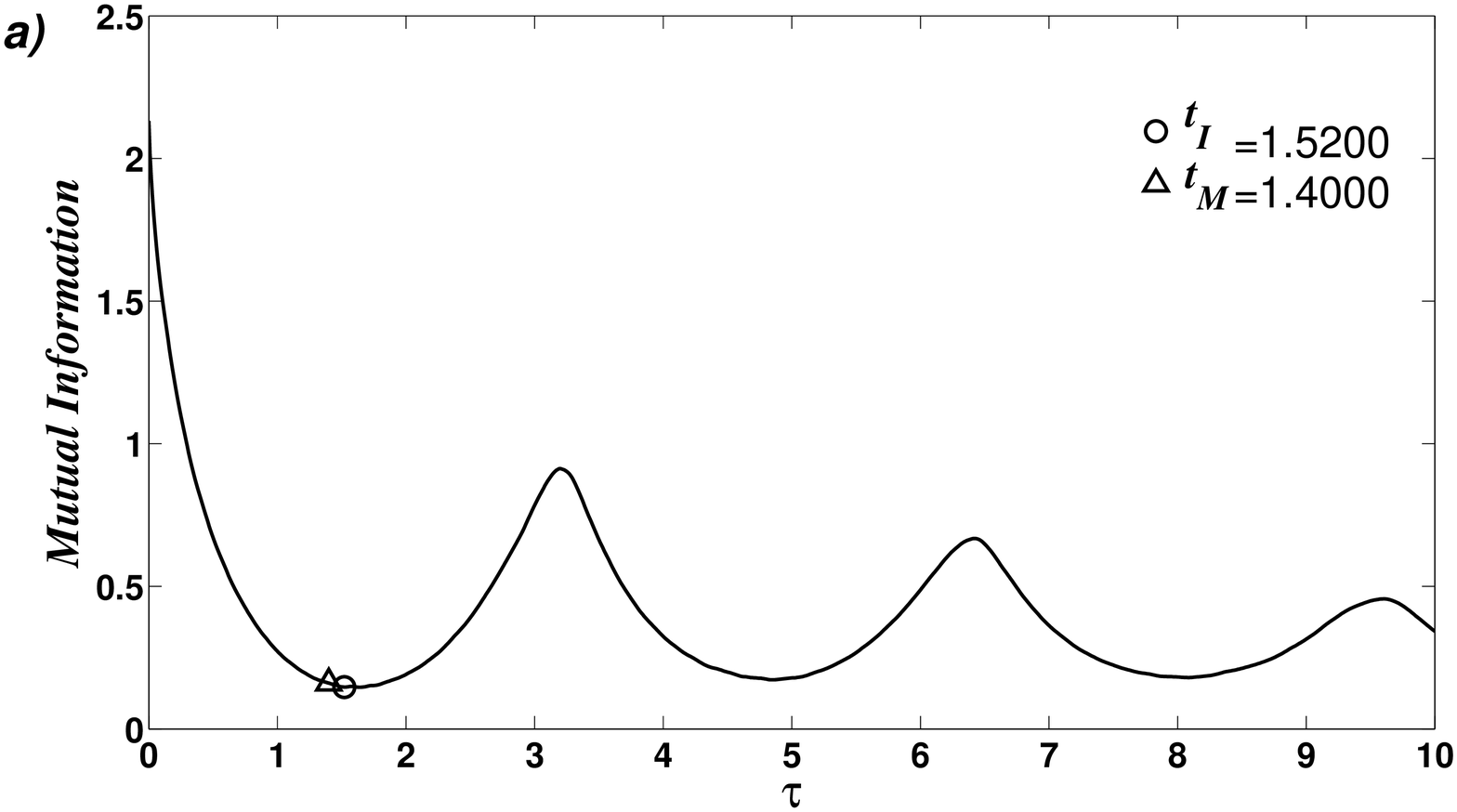}
\includegraphics[ width=0.8\textwidth]{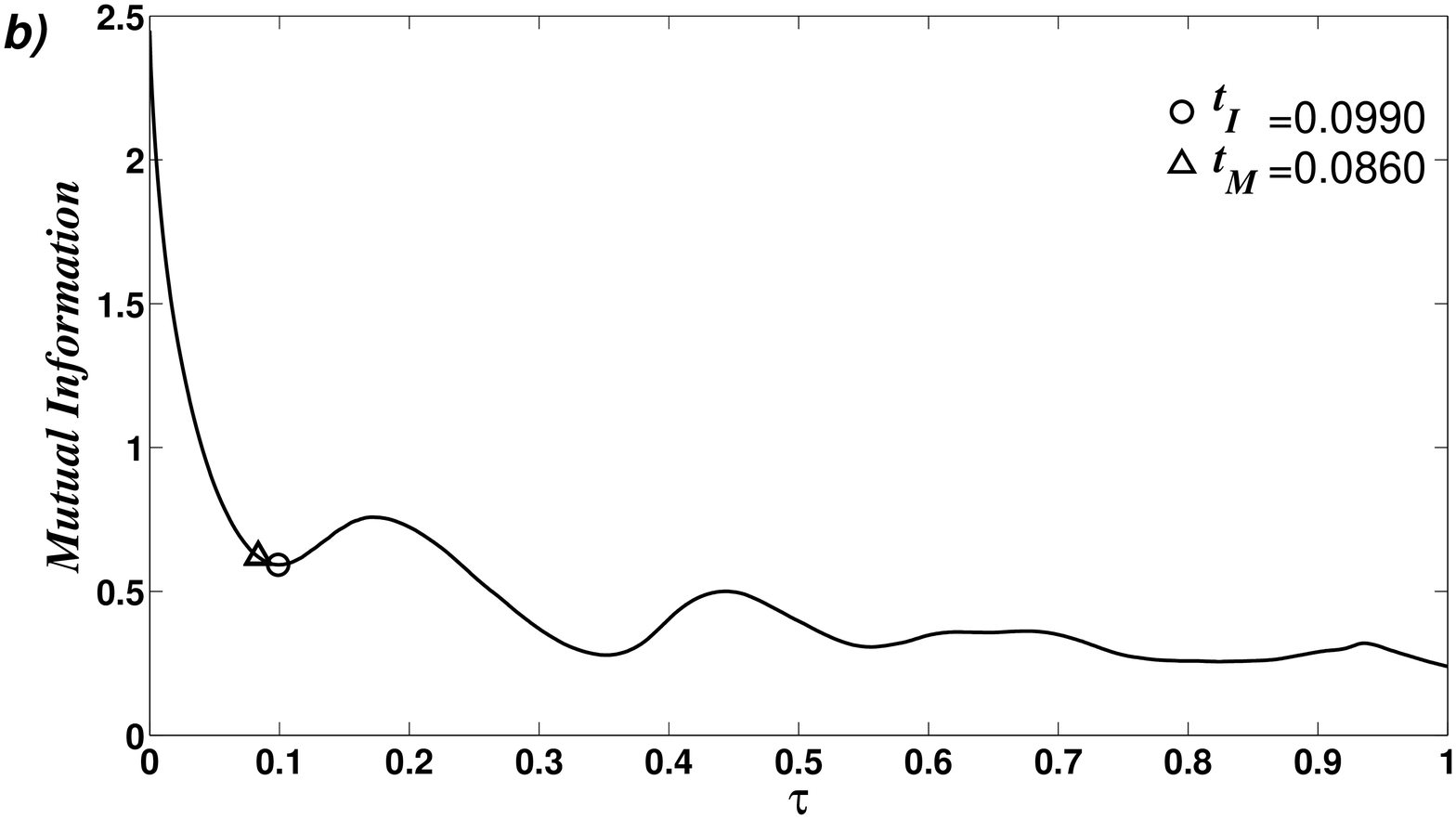}
\includegraphics[ width=0.8\textwidth]{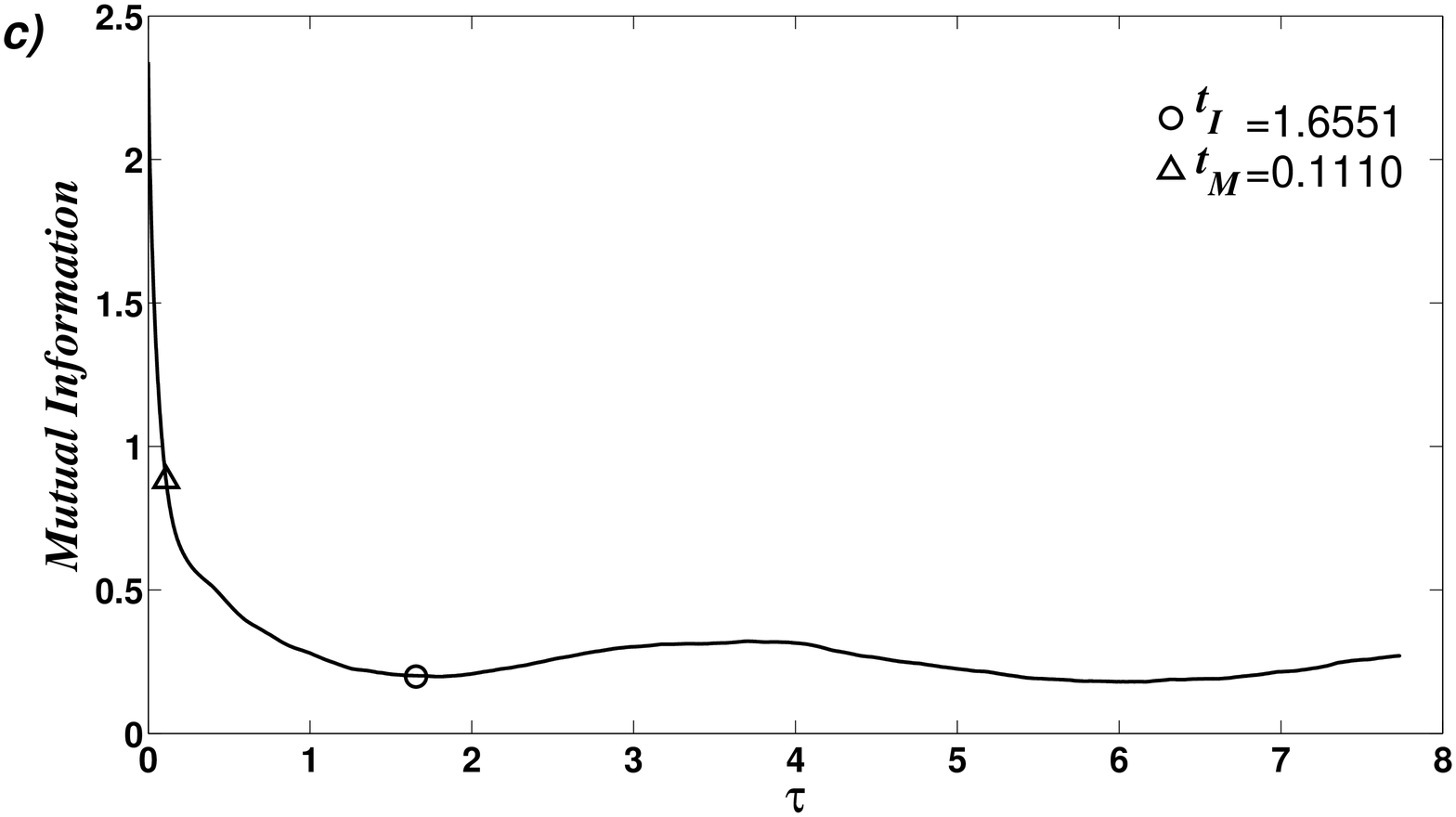}

\caption{ Discrete mutual information $I_m$ for:
{\it (a)\/} Rossler System with  $a=0.45$, $b=2$,  $c=4$.
{\it (b)\/} Lorenz System with $\sigma= 16$, $r=45.92 $, and $b=4$.
{\it (c)\/} ${\mathbf B}_7$ System with $K~=~0.5$, $\alpha~=~7.0$ and $\epsilon~=~0.23$.
In all the cases the time series length
considered had $M = 10^5~data$. The characteristic time induced by
$C^{(BP)}$ ($d=6$), that is $t_M$ and its closest-lying
characteristic time induced by $I_m$  are shown. }
\label{fig:IM}
\end{figure}

%
\begin{figure}
\newpage
\includegraphics[ width=0.8\textwidth]{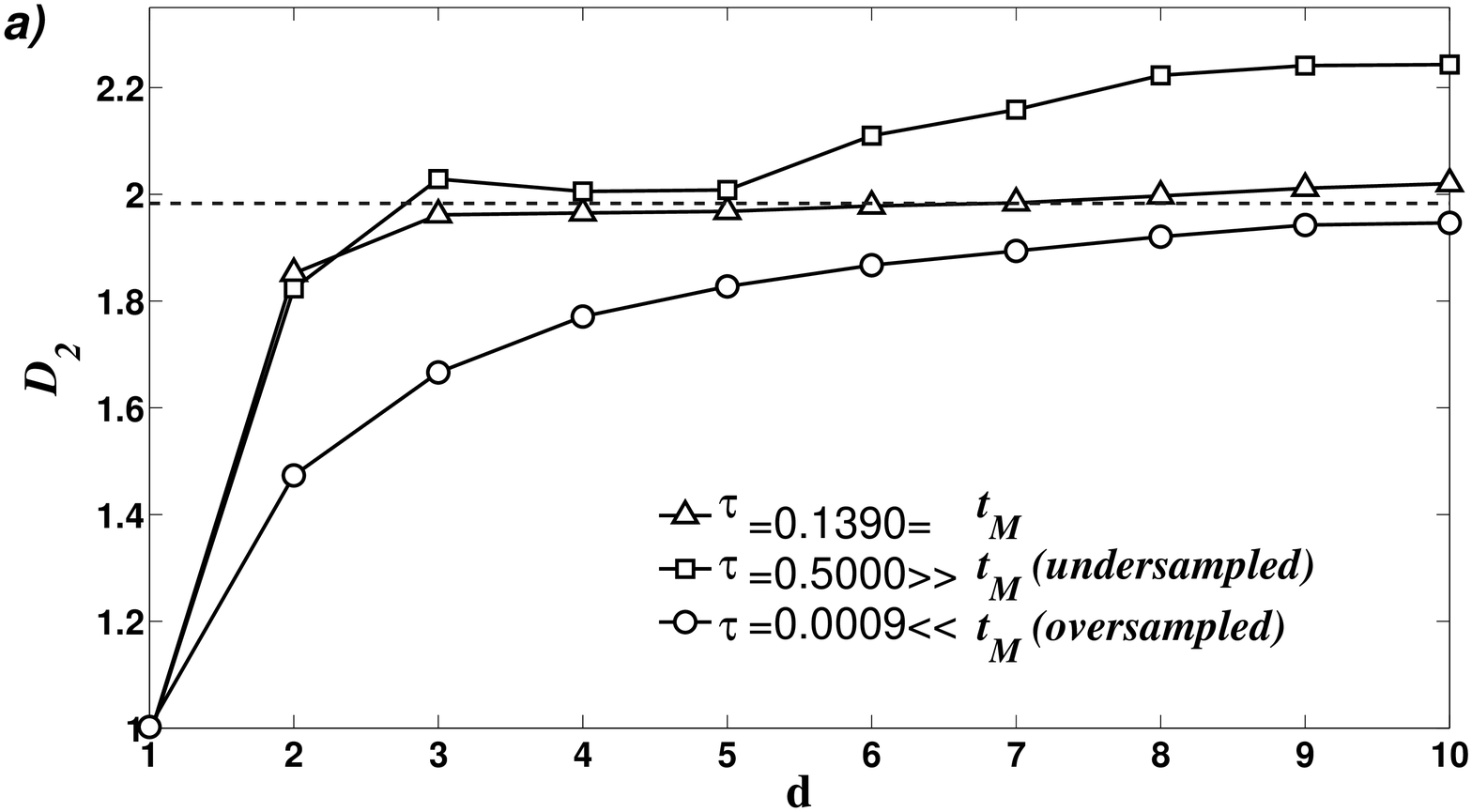}
\includegraphics[ width=0.8\textwidth]{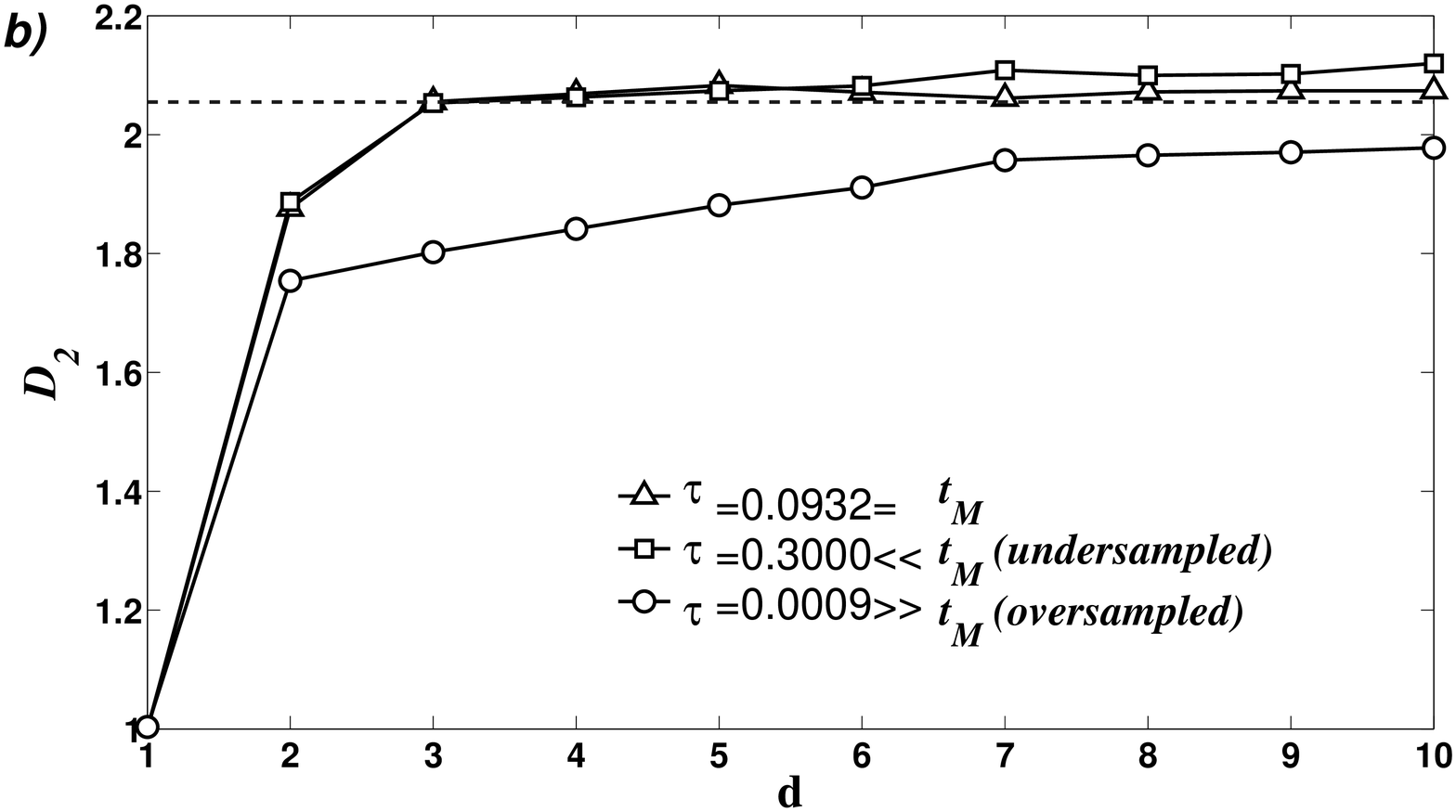}
\includegraphics[ width=0.8\textwidth]{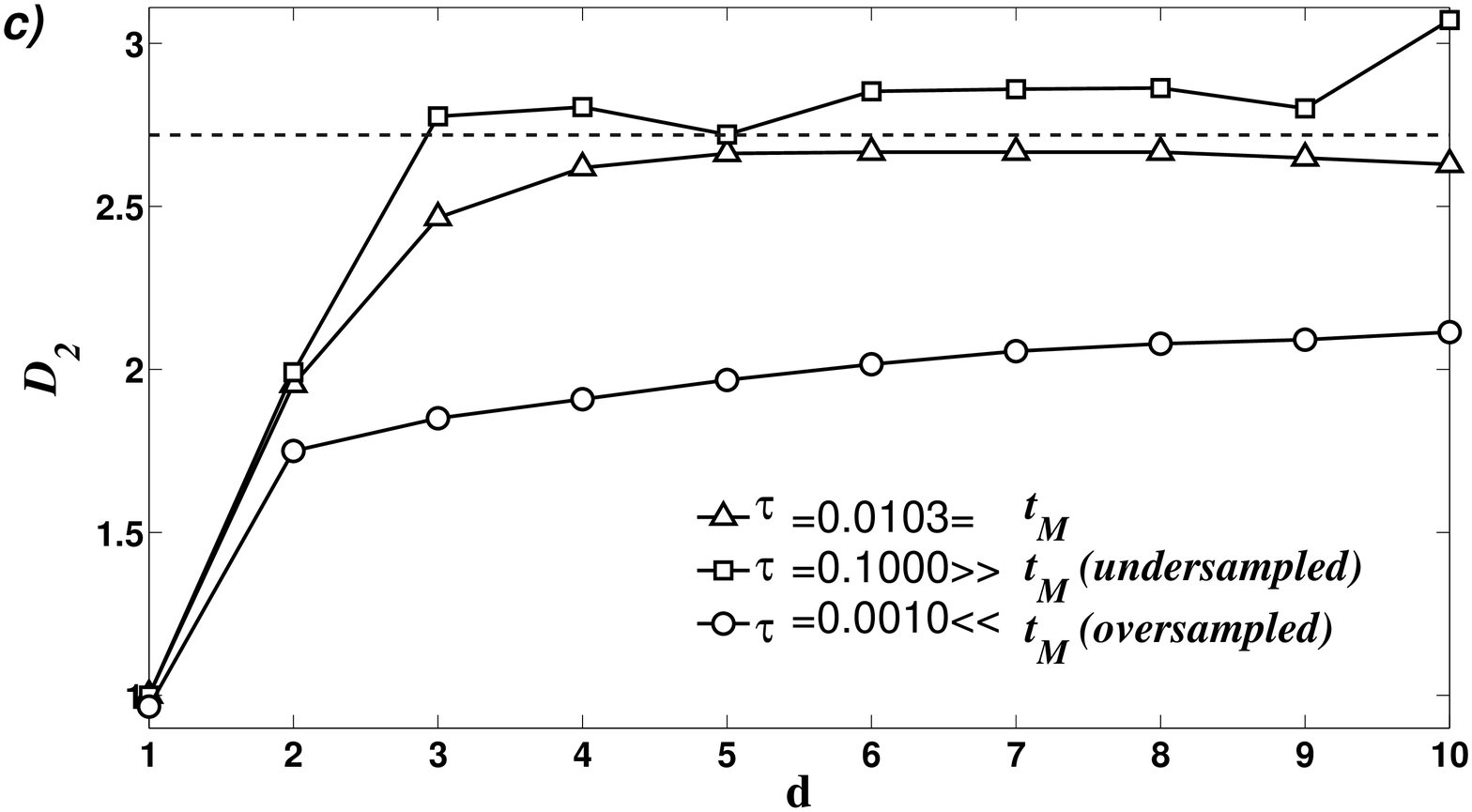}
\caption{ Mean value over realizations of the correlation
dimension  ($<D_2>$)  for sampled $x$ ($y$ and $z$ samplings
produce similar results) as a function of  the embedding dimension $d$:
{\it (a)\/} Rossler System with  $a=0.45$, $b=2$,  $c=4$.
{\it (b)\/} Lorenz System with $\sigma= 16$, $r=45.92 $, and $b=4$.
{\it (c)\/} ${\mathbf B}_7$ System with $K~=~0.5$, $\alpha~=~7.0$, and $\epsilon~=~0.23$.
In all the cases the time
series length  considered had $M= 10^5~data$. For oversampled
series the value of $D_2$ remains below the reference value for
any embedding dimension. For undersampled series all linear and
nonlinear correlations are lost and the system behaves like a
stochastic one. The correlation dimension $D_2$ increases with the
embedding dimension $d$.}
\label{fig:d2A9}
\end{figure}

\newpage
\begin{table}

\caption{ Characteristic Times for coordinates $x$, $y$, $z$  and
their standard deviation (denoted by $\sigma_i$) of the Rossler,
Lorenz and ${\mathbf B}_7$ chaotic attractors.
System's parameters are $a=0.45$, $b=2$,  $c=4$ for Rossler,
$\sigma= 16$, $r=45.92$, and $b=4$ for Lorenz and
$K=0.5$, $\alpha=7.0$, and $\epsilon=0.23$ for ${\mathbf B}_7$.
In all the cases the time series length  considered had $M = 10^5~data$.
$<x,y,z>$ represent the corresponding time obtained averaging  $C^{(BP)}$ over the three coordinates.
Each characteristic time is classified according with the method used to obtain it.%
Numbers in bold face indicate the characteristic time of each
method that best approximates $t_M$ 
(the time corresponding
to the maximum of the Statistical Complexity Measure $C^{(BP)}$ )($d=6$).
}
\begin{center}
\begin{tabular}{|| c || c | c || c | c || c | c || c|| c||}
\hline
\hline
\multicolumn{9}{||c||}{\textbf{Rossler System}} \\
 \hline
time               &$x$      &$\sigma_x$    &$y$      &$\sigma_y$    &$z$      &$\sigma_z$    &$<x, y, z>$   &$t/t_M$ \\
\hline
\multicolumn{9}{||c||}{Time induced by Complexity} \\
\hline
$t_M$              &1.3970   & 0.0067       &1.8090   &0.1473        &1.4050   &0.0053        &1.4000        &1.000 \\
\hline
\multicolumn{9}{||c||}{Times induced by Autocorrelation} \\
\hline
$t_0$              &1.4560   &0.0178        &1.5600   &$<.00001$     &1.7150   &0.0118        &1.6000       &\textbf{1.1429} \\
$t_1$              &3.2070   &0.0157        &3.0820   &0.0063        &3.2350   &0.0108        &3.1600       &2.2571 \\
$t_2$              &0.6380   &0.0079        &1.3520   &0.0103        &1.5190   &0.0032        &1.0000       &0.7143 \\
$t_{1/e}$          &0.8480   &0.0157        &1.1700   &0.0063        &1.2930   &0.0108        &1.1200       &0.8000 \\
\hline
\multicolumn{9}{||c||}{Time induced by Mutual Information} \\
\hline
$t_I$              &1.6240   &0.0534        &1.4530   &0.0874        &1.5600   &0.0337        &1.5200       &\textbf{1.0857} \\
\hline
\multicolumn{9}{||c||}{Times induced by Nyquist-Shannon} \\
\hline
$t_{NS}^{(0.85)}$  &0.8795   &--            &1.4645   &--            &2.0956    &--           &1.4799       &\textbf{1.0571} \\
$t_{NS}^{(0.90)}$  &0.7292   &--            &1.1418   &--            &1.6661    &--           &1.1791       & 0.8422 \\
$t_{NS}^{(0.95)}$  &0.5691   &--            &0.7447   &--            &0.9142    &--           &0.7427       &0.5305 \\
\hline
\hline
\end{tabular}
\label{tab:chartimes2}
\end{center}
\end{table}

%
%
\setcounter{table}{0}
\begin{table}
\caption{Continuation.}
\begin{center}
\begin{tabular}{|| c || c | c || c | c || c | c || c|| c||}
\hline
\hline
\multicolumn{9}{||c||}{\textbf{Lorenz System}} \\
\hline
time               & $x$     &$\sigma_x$    &$y$      &$\sigma_y$    &$z$       &$\sigma_z$   &$<x, y, z>$   &$t/t_M$\\
\hline
\multicolumn{9}{||c||}{Time induced by Complexity} \\
\hline
$t_M$              &0.0932   &0.0004        &0.0722   &0.0010        &0.0853    &0.0007       & 0.0860       &1.0000\\
\hline
\multicolumn{9}{||c||}{Times induced by Autocorrelation} \\
\hline
$t_0$              &2.3805   &1.0482        &1.9424   &1.1457        &0.1217    &0.0018       &0.2130        &2.4767\\
$t_1$              &0.4226   &0.0308        &0.3979   &0.0342        &0.2334    &0.0007       &0.2580        &3.0000 \\
$t_2$              &0.1012   &0.0006        &0.0760   &0.0001        &0.0874    &0.0005       &0.0860        &\textbf{1.000} \\
$t_{1/e}$          &0.2078   &0.0308        &0.1478   &0.0342        &0.0883    &0.0007       &0.1230        &14.3023 \\
\hline
\multicolumn{9}{||c||}{Time induced by Average Mutual Information} \\
\hline
$t_I$              &0.1048   &0.0006        &0.0970   &0.0012        &0.0941    &0.0028       &0.0990        &\textbf{1.1512} \\
\hline
\multicolumn{9}{||c||}{Times induced by Nyquist-Shannon} \\
\hline
$t_{NS}^{(0.85)}$  &0.1383   &--            &0.1082   &--            &0.0884    &--           &0.1116        &1.2981 \\
$t_{NS}^{(0.90)}$  &0.1138   &--            &0.0880   &--            &0.0670    &--           &0.0896        &\textbf{1.0419}\\
$t_{NS}^{(0.95)}$  &0.0756   &--            &0.0639   &--            &0.0084    &--           &0.0493        &0.5733 \\
\hline
\hline
\end{tabular}
\label{tab:chartimes2-cont1}
\end{center}
\end{table}
%

%
%
\setcounter{table}{0}
\begin{table}
\caption{Continuation.}
\begin{center}
\begin{tabular}{|| c || c | c || c | c || c | c || c|| c||}
\hline
\hline
\multicolumn{9}{||c||}{\textbf{${\mathbf B}_7$ System}} \\
\hline
time                &$x$       &$\sigma_x$    &$y$      &$\sigma_y$    &$z$      &$\sigma_z$    &$<x, y, z>$   &$t/t_M$\\
\hline
\multicolumn{9}{||c||}{Time induced by Complexity} \\
\hline
$t_M$              &0.1030     &$<0.001$      &0.0930    &$<0.001$     &0.9540   &$<0.001$      &0.3833        &1.0000\\
\hline
\multicolumn{9}{||c||}{Times induced by Autocorrelation} \\
\hline
$t_0$              &0.5459    &$0.3775$       &0.5635    &$0.4363$     &1.9985   &0.0258        &1.9860        &5.1808\\
$t_1$              &0.4509    &$0.1332$       &0.3915    &$0.0957$     &3.9141   &0.1161        & 3.6870       &9.6182\\
$t_2$              &0.1301    &$0.0129$       &0.1301    &$0.0127$     &1.5631   &0.0227        &0.1280        &0.3339\\
$t_{1/e}$          &0.1813    &$0.1332$       &0.1747    &$0.0957$     &1.4504   &0.1161        &0.2830        &\textbf{0.7382}\\
\hline
\multicolumn{9}{||c||}{Time induced by Average Mutual Information} \\
\hline
$t_I$              &1.5289    &0.2240         &1.5951    &0.2212       &1.7905   &0.1041        &1.6551        &\textbf{4.3176} \\
\hline
\multicolumn{9}{||c||}{Times induced by Nyquist-Shannon} \\
\hline
$t_{NS}^{(0.85)}$  &0.1901    &--             &0.1968    &--           &0.5319   &--            &0.3062        &\textbf{0.7987}\\
$t_{NS}^{(0.90)}$  &0.1683    &--             &0.1736    &--           &0.0731   &--            &0.1383        &0.3607\\
$t_{NS}^{(0.95)}$  &0.0539    &--             &0.1441    &--           &0.0076   &--            &0.06853       &0.1787 \\
\hline
\hline
\end{tabular}
\label{tab:chartimes2-cont2}
\end{center}
\end{table}


\begin{thebibliography}{99}
%
%
\bibitem{Crutchfield1989}
J. P. Crutchfield, K. Young.
Inferring statistical complexity.
Phys. Rev. Lett. 63 (1989) 105--108.

\bibitem{Wackerbauer1994}
R. Wackerbauer, A. Witt, H. Atmanspacher, J. Kurths, H. Scheingraber.
A comparative classification of complexity measures.
Chaos, Solitons \& Fractals, 4 (1994) 133--173.

\bibitem{Lopez1995}
R. L\'opez-Ruiz, H. L. Mancini, X. Calbet.
A statistical measure of complexity.
Phys. Lett. A 209 (1995) 321--326.

\bibitem{Martin2003}
M. T. Mart\'{\i}n, A. Plastino, and O. A. Rosso.
Statistical complexity and disequilibrium.
Phys. Lett. A 11 (2003) 126--132.

\bibitem{Lamberti2004}
P. W. Lamberti, M. T. Mart\'{\i}n, A. Plastino, and O. A. Rosso.
Intensive entropic non-triviality measure.
Physica A 334 (2004) 119--131.

\bibitem{Rosso2007A}
O. A. Rosso, H. A. Larrondo, M. T. Mart\'{\i}n, A. Plastino, M. A. Fuentes.
Distinguishing noise from chaos.
Phys. Rev. Lett. 99 (2007) 154102.

\bibitem{Rosso2007B}
O. A. Rosso, L. Zunino, D. G. P\'erez, A. Figliola, H. A. Larrondo, M. Garavaglia,
M. T. Mart\'{\i}n,  A. Plastino.
Extracting features of gaussian selfsimilar stochastic processes via the Bandt \& Pompe approach.
Phys. Rev. E, 76 (2007) 061114.

\bibitem{DeMicco2008}
L. De Micco, C. M. Gonz\'alez, H. A. Larrondo, M. T.  Mart\'{\i}n, A. Plastino, O. A. Rosso.
Randomizing nonlinear maps via symbolic dynamics.
Physica A 387 (2008) 3373--3383.

\bibitem{DeMicco2009}
L. De Micco, H. Larrondo, A. Plastino, O. A. Rosso.
Quantifiers for stochasticity of chaotic pseudo random number generators.
Phil. Trans.  Royal Soc. A  367  (2009) 3281--3296.

\bibitem{Nyquist1928}
H. Nyquist.
Certain topics in telegraph transmission theory.
Trans. AIEE 47 (1928) 617--644.

\bibitem{Shannon1949}
C. E. Shannon.
Communication in the presence of noise.
Proc. Institute of Radio Engineers 37 (1949) 10--21.

\bibitem{Takens1981}
F. Takens.
Detecting strange attractors in turbulence.
Lecture Notes in Mathematics 898 (1981) 366--381.

\bibitem{Hegger1999}
R. Hegger, H. Kantz, and T. Schreiber.
Practical implementation of nonlinear time series methods: The tisean package.
Chaos 9 (1999) 413--435.

\bibitem{Pompe2002}
C. Bandt, B. Pompe.
Permutation entropy: a natural complexity measure for time series.
Phys. Rev. Lett. 88 (2002) 174102.

\bibitem{Pearl2009}
J. Pearl.
Causality: models, reasoning, and inference.
Cambridge University Press, 2009.

\bibitem{Fraser1986}
A. M. Fraser, H. L. Swinney.
Independent coordinates for strange attractors from mutual information.
Phys. Rev. A 33 (1986) 1134--1140.

\bibitem{Crutchfield1987}
J. P. Crutchfield, B. McNamara.
Equations of motion from a data series.
Complex Systems 1 (1989) 417--452.

\bibitem{Kantz1999}
H. Kantz, T. Shreiber.
Nonlinear Time Series Analysis.
Cambridge University Press, 1999.

\bibitem{Chlouverakis2005}
K. E. Chlouverakis and J. C. Sprott.
A comparison of correlation and lyapunov dimensions.
Physica D 200 (2005) 156--164.

\bibitem{Soriano2011}
M. C. Soriano, L. Zunino, O. A. Rosso, I. Fischer, C. R. Mirasso.
Time scales of chaotic semiconductor laser with optical feedback under the lens of permutation information analysis.
IEEE J. Quantum Electronics 47 (2011) 252--261.


\bibitem{Sauer1993}
T. Sauer, J. A. Yorke, M. Casdagli.
Embedology.
Journal of Statistical Physics 65 (1991) 579--616.

\bibitem{Casdagli1991}
M. Casdagli, S. Eubank, J. D. Farmer, J. Gibson.
State space reconstruction in the presence of noise.
Physica D  51 (1991) 52--98.

\bibitem{Gibson1992}
J. Gibson, J. D. Farmer, M. Casdagli, S. Eubank.
An analytical approach to practical state space reconstruction.
Physica D  57 (1992) 1--30.

\bibitem{Abarbanel1996}
H. D. I. Abarbanel.
Analysis of observed chaotic data.
Springer-Verlag, New York, 1996.

\bibitem{Martin2006}
M. T. Mart\'{\i}n, A. Plastino, O. A. Rosso.
Generalized statistical complexity measures: geometrical and analytical properties.
Physica A 369 (2006) 439--462.

\bibitem{Rosso2009C}
O. A. Rosso, H. Craig, P. Moscato,
Shakespeare and other English renaissance authors as characterized by Information Theory complexity quantifiers.
Physica A 388 (2009) 916--926

\bibitem{Mischaikow1999}
K. Mischaikow, M. Mrozek, J. Reiss, A. Szymczak.
Construction of symbolic dynamics from experimental time series.
Phys. Rev. Lett., 82 (1999) 1114--1147.

\bibitem{Powell1979}
G. E. Powell, I. C. Percival.
A spectral entropy method for distinguishing regular and irregular motion of hamiltonian systems.
J. Phys. A: Math. Gen. 12 (1979) 2053--2071.

\bibitem{Blanco1998}
S. Blanco, A. Figliola, R. Quian Quiroga, O. A. Rosso, E. Serrano.
Time-frequency analysis of electroencephalogram series (III): Wavelet packets and information cost function.
Phys. Rev. E 57 (1998) 932--940.

\bibitem{Rosso2001}
O. A. Rosso, S. Blanco, J. Jordanova, V. Kolev, A. Figliola, M. Sch\"urmann, E. Ba\c{s}sar.
Wavelet entropy: a new tool for analysis of short duration brain electrical signals.
Journal of Neuroscience Methods 105 (2001) 65--75.

\bibitem{Ebeling2001}
W. Ebeling, R. Steuer.
Partition-based entropies of deterministic and stochastic maps.
Stochastics and Dynamics, 1 (2001) 1--17.

\bibitem{Amigo2007}
J. M. Amig\'o, L. Kocarev, I. Tomovski.
Discrete entropy.
Physica D 228 (2007) 77--85.

\bibitem{Keller2005}
K. Keller, M. Sinn.
Ordinal analysis of time series.
Physica A  356 (2005) 114--120.

\bibitem{Rosso2009A}
O. A. Rosso, C. Masoller.
Detecting and quantifying stochastic and coherence resonances via information theory complexity measurements.
Phys. Rev. E 79 (2009) 040106(R).

\bibitem{Rosso2009B}
O. A. Rosso, C. Masoller.
Detecting and quantifying temporal correlations in stochastic resonance via information theory measures.
European Phys. Journal B, 69 (2009) 37--43.

\bibitem{Larrondo2005}
H. A. Larrondo, C. M. Gonz\'alez, M. T. Mart\'{\i}n, A. Plastino, O. A. Rosso.
Intensive statistical complexity measure of pseudorandom number generators.
Physica A  356 (2005) 133--138.

\bibitem{Larrondo2006}
H. A. Larrondo, M. T. Mart\'{\i}n, C.M. Gonz\'alez, A. Plastino, O. A. Rosso.
Random number generators and causality.
Phys. Lett. A 352 (2006) 421--425.

\bibitem{Kowalski2007}
A. M. Kowalski, M. T. Mart\'{\i}n, A. Plastino, O. A. Rosso.
Bandt-Pompe approach to the classical-quantum transition.
Physica D  233 (2007) 21--31.

\bibitem{Zunino2007}
L. Zunino, D. G. P\'erez, M. T. Mart\'{\i}n, A. Plastino, M. Garavaglia, O. A. Rosso.
Characterization of gaussian self-similar stochastic processes using wavelet-based informational tools.
Phys. Rev. E  75 (2007) 021115.

\bibitem{Rosso2008}
O. A. Rosso, R. Vicente, C. R. Mirasso.
Encryption test of pseudo-aleatory messages embedded on chaotic laser signals: an information theory approach.
Phys. Lett. A 372 (2008) 1018--1023.

\bibitem{Zunino2008A}
L. Zunino, D. G. P\'erez, M. T. Mart\'{\i}n, M. Garavaglia, A. Plastino, O. A. Rosso.
Permutation entropy of fractional brownian motion and fractional gaussian noise.
Physics Letters A  372 (2008) 4768--4774.

\bibitem{Zunino2008B}
L. Zunino, D. G. P\'erez, A. Kowalski, M. T. Mart\'{\i}n, M. Garavaglia, A. Plastino, O. A. Rosso.
Fractional Brownian motion, fractional Gaussian noise, and Tsallis permutation entropy.
Physica A  387 (2008) 6057--6068

\bibitem{Zunino2009}
L. Zunino, M. Zanin, B. M. Tabak, D. Perez, and O. A. Rosso.
Forbidden patterns, permutation entropy and stock market inefficiency.
Physica A 388 (2009) 2854--2864

\bibitem{Zunino2010b}
L. Zunino, M. Zanin, B. M. Tabak, D. G. P\'erez, O. A. Rosso.
Complexity-entropy causality plane: a useful approach to quantify the stock market inefficiency.
Physica A 389 (2010) 1891--1901.

\bibitem{Rosso2010A}
O. A. Rosso, L. De Micco, H. Larrondo, M. T. Mart\'{\i}n, A. Plastino.
Generalized statistical complexity measure.
Int. J. Bif. and Chaos 20 (2010) 775--785.

\bibitem{Rosso2010B}
O. A. Rosso, L. De Micco, A. Plastino, H. A. Larrondo.
Info-quantifiers' map-characterization revisited.
Physica A 389 (2010) 4604--4612.

\bibitem{Saco2010}
P. M. Saco, L. C. Carpi, A. Figliola, E. Serrano, and O. A. Rosso.
Entropy Analysis of the Dynamics of EL Ni\~no/Southern Oscillation during the Holocene.
Physica A 389 (2010) 5022--5027.

\bibitem{Keller2003}
K. Keller, H. Lauffer.
Symbolic analysis of high-dimensional time series.
Int. J. Bifurcation and Chaos 13 (2003) 2657--2668.

\bibitem{Press1995}
W. H. Press, S. A. Teikolsky, W. T. Vetterling, B. P. Flannery.
Numerical Recipes in C.
Cambridge University Press, 1995.

\bibitem{Setti2002}
G. Setti, G. Mazzini, R. Rovatti, and S. Callegari.
Statistical modeling of discrete-time chaotic processes: Basic finite-dimensional tools and applications.
Proceedings of the IEEE 90 (2002) 662--689.

\bibitem{Callegari2003A}
S. Callegari, R. Rovatti, and G. Setti.
Chaos-based fm signals: application and implementation issues.
IEEE Transactions on Circuits and Systems I: Fundamental Theory and Applications 50 (2003) 1141--1147.

\bibitem{Sprott2005}
J. C. Sprott.
Lyapunov exponent and dimension of the lorenz attractor.
http://sprott.physics.wisc.edu/chaos/lorenzle.htm, 2005.

\bibitem{Sprott2001}
J. C. Sprott, C. Rowlands.
Improved correlation dimension calculation.
Int. J. Bifurcation and Chaos 11 (2001) 1865--1880.

\bibitem{Zunino2010}
L. Zunino, M. C. Soriano,  I. Fischer, O. A. Rosso, C. R. Mirasso.
Permutation-information-theory approach to unveil delay dynamics from time-series analysis.
Phys. Rev. E 82 (2010) 046212.



\end{thebibliography}
\end{document}